\documentclass[fleqn,usenatbib]{mnras}

\usepackage{newtxtext,newtxmath}

\usepackage[T1]{fontenc}
\usepackage{ae,aecompl}

\usepackage[dvipsnames]{xcolor}
\usepackage{graphicx}
\usepackage{amsmath}
\usepackage{amssymb}
\usepackage{placeins}
\usepackage{color}
\usepackage{microtype}
\usepackage{booktabs}
\usepackage{threeparttable}
\usepackage{tabularx}

\newcommand{\given}{\,|\,}
\newcommand{\dd}{\mathbf{D}}

\newcommand{\pars}{\theta}
\newcommand{\hypers}{\alpha}

\newcommand{\likelihood}{{\mathcal L}}

\title[Magnetar Giant Flares]{Magnetar giant flare high-energy emission}

\author[C. Elenbaas et al.]{C. Elenbaas,$^{1}$\thanks{E-mail: cpc.elenbaas@gmail.com} D. Huppenkothen,$^{2,3}$  C. Omand,$^{4,5}$  A.L. Watts,$^{1}$  E. Bissaldi,$^{6,7}$ 
\newauthor I. Caiazzo$^{4}$ and J. Heyl$^{4}$ \\
$^{1}$Anton Pannekoek Institute for Astronomy, University of Amsterdam, Science Park 904, 1098 XH Amsterdam, the Netherlands
\\
$^{2}$Center for Cosmology and Particle Physics, Department of Physics, New York University, 4 Washington Place, New York, NY 10003, USA
\\
$^{3}$Center for Data Science, New York University, 65 5th Avenue, 7th Floor, New York, NY 10003, USA
\\
$^{4}$Department of Physics and Astronomy, University of British Columbia, 6224 Agricultural Road, Vancouver, British Columbia V6T 1Z1, Canada
\\
$^{5}$Department of Physics, Graduate School of Science, The University of Tokyo, 7-3-1 Hongo, Bunkyo-ku, Tokyo 113-8656, Japan
\\
$^{6}$Dipartimento Interateneo di Fisica, Politecnico di Bari, Via E.Orabona 4, 70125 Bari, Italy
\\
$^{7}$INFN - Sezione di Bari, Via E.Orabona 4, I-70125 Bari, Italy  
\\
}

\pubyear{2017}

\begin{document}
\label{firstpage}
\pagerange{\pageref{firstpage}--\pageref{lastpage}}

\maketitle

\begin{abstract}
High energy ($> 250$ keV) emission has been detected persisting for several tens of seconds after the initial spike of magnetar giant flares.  It has been conjectured that this emission might arise via inverse Compton scattering in a highly extended corona generated by super-Eddington outflows high up in the magnetosphere. In this paper we undertake a detailed examination of this model.  We investigate the properties of the required scatterers, and whether the mechanism is consistent with the degree of pulsed emission observed in the tail of the giant flare. We conclude that the mechanism is consistent with current data, although the origin of the scattering population remains an open question.  We propose an alternative picture in which the emission is closer to that star and is dominated by synchrotron radiation. The \emph{RHESSI} observations of the December 2004 flare modestly favor this latter picture. We assess the prospects for the Fermi Gamma-Ray Space Telescope to detect and characterize a similar high energy component in a future giant flare. Such a detection should help to resolve some of the outstanding issues.  
\end{abstract}


\section{Introduction}		

Magnetars are Neutron Stars (NSs) with magnetic fields at the extreme high end of the distribution, in the range $10^{13} - 10^{15}$ G.  One of their defining characteristics, indeed the one that led to their discovery \citep{Mazets1979a,Mazets1981}, is the repeated emission of bursts of hard X-rays and $\gamma$-rays.  The bursts are assumed to be powered by magnetic field decay \citep{Thompson1995}, yet many details of the burst trigger and emission process remain unsolved \citep[see][for a recent review]{Turolla2015}. 

Bursts come in a range of fluences, but the brightest are the rare Giant Flares (GFs) with energies in the range $10^{44}-10^{46}$ erg (if the emission is isotropic).  Only three have been observed over the last 38 years, each from a different magnetar.  Their properties are very similar:  a very bright initial spike lasting $\sim 0.1 - 1$ s followed by a decaying tail lasting several hundred seconds, strongly pulsed at the few second spin period of the star \citep{Mazets1979a,Hurley1999,Hurley2005}.  

The initial $\gamma$-ray spike is very hard, with emission being detected up to a few MeV \citep{Mazets1979a,Hurley1999}.  It is thought to originate in particle acceleration as the evolving magnetic field reaches a tipping point and undergoes explosive reconfiguration or reconnection. Rapid acceleration of electrons in a strong curved magnetic field leads naturally to a cascade of $\gamma$-rays and pair creation \citep{Sturrock1989}. The existence of radio afterglows points to the simultaneous ejection of a plasmoid of relativistic particles and magnetic fields that then energizes a pre-existing shell of surrounding material \citep{Frail1999,Cameron2005,Granot2006}. 

Many of the electron-positron pairs from the initial event become trapped in closed field regions, with optical thickness high enough to trap the photons, leading to rapid thermalization \citep{Thompson1995}.  This hot pair plasma fireball cools and contracts relatively slowly, generating the long tail. The emission is strongly beamed by super-Eddington mildly relativistic outflows driven by the radiation escaping from the fireball \citep{Thompson1995,Thompson2001,vanPutten2013,vanPutten2016}.  Although the temperature in the core of the fireball is thought to be $\sim 100$ keV, the emerging radiation has a photospheric temperature $\sim 10-30$ keV. 

Following the GF from the magnetar SGR 1806-20 in December 2004, \citet{Boggs2007} reported the detection of two hard nonthermal components in the spectrum after the initial spike. One power law component extends up to $\gtrsim 250$ keV and persists throughout the pulsed tail.  However there is also a second power law component that extends up to 17 MeV (the limit of the \emph{RHESSI} instrument) with no sign of a cutoff. \citet{Frederiks2007} also report the observation of a high energy power law component extending up to the sensitivity limit of the Konus-Wind detector, i.e. 10 MeV. A hard tail ($\sim 300-650$ keV) was observed by \citet{Feroci1999} for the 1998 August 27 GF of SGR 1900+14, where the upper limit was constrained by the sensitivity of the Gamma-Ray Burst Monitor (GRBM) on board the \emph{BeppoSAX} spacecraft.

The nonthermal component at the lower end of the energy spectrum is thought to emerge from electron cyclotron scattering in an extended corona ($\sim5-10\,R_{\rm NS}$, where $R_{\rm NS}$ is the radius of the NS), where output photon energies in excess of 100 keV are attainable \citep{Thompson2002,Boggs2007}. The hard energy component that reaches tens of MeV however is conjectured by \citet{Feroci2001} and \citet{Boggs2007} to originate in a \emph{highly} extended corona, generated by super-Eddington outflows, high up in the magnetosphere ($\sim100\,R_{\rm NS}$). At these altitudes the magnetic energy density has become less than the energy density of the X-ray emission, i.e. $u_B<u_{\rm X}$, and the charges cool through inverse Compton scattering, rather than synchrotron emission. The hard nonthermal tail is thus argued to be the result of inverse Compton (IC) scattering of the emission onto the charges that make up this additional extended corona\footnote{The smooth $\sim40$ s decay that followed the initial spike of the GF from SGR 1900+14 on 1998 August 27, is argued to have been the signature of a pair corona, produced by post burst onset seismic activity, that surrounded the fireball ($\sim10\,R_{\rm NS}$) and gradually evaporated \citep{Thompson2001,Feroci2001}. This corona proceeded to Compton scatter O-mode photons seeping out at the base of the fireball hardening the emergent emission. Note however that this corona is dissimilar from the highly extended corona ($\sim100\,R_{\rm NS}$), invoked to explain the second hard component ($\sim$ 0.4 - 17 MeV), that upscatters photons at the outer edges of the X-ray jets.}. 

Note that during quiescence a nonthermal power law component is also observed in the hard X-ray range ($> 20$ keV) of various magnetars: 4U 0142+614, 1RXS J1708-4009, 1E 1841-045, 1E 2259+586, SGR 1806- 20, and SGR 1900+14 \citep{Kuiper2004,Kuiper2006,Mereghetti2005a,Molkov2005,Gotz2006,Enoto2011,Turolla2015}. The non-detection of $\gamma$-ray emission by Comptel and Fermi \emph{LAT} from quiescent magnetars, suggest a spectral cutoff of this component at $\sim$ few hundred keV \citep{denHartog2006b,Kuiper2006,SasmazMus2010}. Various mechanisms have been invoked to explain the hard quiescent emission. \citet{Thompson2005} consider two scenarios: (i) thermal Bremsstrahlung through crustal heating by return currents and (ii) synchrotron radiation (peaking at $\sim1.5$ MeV) from energetic pairs created by Comptonized X-ray photons on accelerated positrons in the magnetosphere (at $\sim10\,R_{\rm NS}$). Another possibility is resonant cyclotron scattering (RCS) of a thermal seed photon population \citep{Baring2005,Baring2007}. This scenario has been studied in detail with relativistic full QED 3D Monte Carlo codes that model the hard component \citep{Nobili2008a,Nobili2008b,Zane2011a}. 

The latter are limited by their use of a self-similar twisted magnetic dipole and simple charge velocity distribution. Phase resolved spectroscopy of AXPs 1RXS 4U 0142+61 and J1708-4009 suggest however the presence of more complex magnetic field configurations \citep{denHartog2008a,denHartog2008b}. The effects of higher multipolar and locally twisted configurations on the emergent high-energy spectra, were investigated by \citet{Pavan2009} and \citet{Vigano2012} respectively. \citet{Beloborodov2012} explored the emergent spectrum from a more physically motivated model for the charge velocity distribution on a locally twisted magnetic bundle.

In quiescence the non-thermal flux is typically $\sim 10^{-12}$ ergs cm$^{-2}$ s$^{-1}$, and the cooling time for RCS is 0.1~ms, about a light-crossing time.  During the burst the non-thermal flux above 1~MeV is $10^6$ times larger, and the non-thermal flux continues nearly $\sim100$ s.  So although the RCS mechanism could account for the instantaneous high-energy spectrum during the burst, the energy content of the quiescent population of high-energy electrons falls short by 12 orders of magnitude.

The purpose of this paper is to examine in detail the mechanism that could generate the highest energy MeV emission, particularly inverse Compton scattering in an extended pair corona high up in the magnetosphere. We consider the properties of the scatterers that would be required to generate the highest energy emission. We contrast these results with those from a model where synchrotron emission dominates the high-energy radiation. We also consider whether these mechanisms are consistent with the observed high-energy spectrum and the degree to which the emission during the tail is pulsed.   Determining amplitudes and upper limits on periodic signals during magnetar bursts is complicated by the fact that the signal has a strong red noise component, as well as the changes in source brightness relative to the sky background imposed by the spectral dependence.  We build a Bayesian model to take both effects into account in our analysis.

We also examine the prospects for future detection of high energy components during GFs. The facility that is best suited to do this is NASA's Fermi Gamma-ray Space Telescope (\emph{FGST} or Fermi) that has been operational since 2008 and will be until at least 2018. Fermi has two instruments: the \emph{Gamma-Ray Burst Monitor} (\emph{GBM}), which provides near full coverage of the unocculted sky in the $\sim0.01-38$ MeV energy band \citep{Meegan2009} and has been a very prolific burst detector \citep{Collazzi2015,Bhat2016}; and the \emph{Large Area Telescope} (\emph{LAT}) provides high energy coverage with a field of view (FoV) $\sim2.4$ sr in the $\sim80-10^4$ MeV energy band \citep{Atwood2009}. The latter has been used to study the $\gamma$-ray upper limits on magnetars \citep{Abdo2010,Li2016}. 

\section{Observations and Analysis}

\subsection{Giant flare high-energy emission}

The Reuven Ramaty High Energy Solar Spectroscopic Imager \emph{RHESSI} is a solar telescope dedicated to the exploration of energetic transient phenomena, such as explosive particle acceleration during solar flares, over a broad energy bandpass (3 keV $-$ 17 MeV) and at a native time resolution of 1 binary $\upmu$s ($2^{-20}$ s $\simeq9.5\times10^{-7}$ s). The principal instrument comprises an array of 9 coaxial high-purity germanium detectors (each with 1 front and 2 rear segments; the former records X-rays up to $\sim100$ keV and the latter detect photons with energies $\gtrsim100$ keV), which are unshielded and may record photons from off-axis sources \citep{Smith2002}. \emph{RHESSI} employs mobile attenuators to inhibit incoming flux from high intensity flares as to prevent saturation due to excessive count rates; their motion however can lead to significant instrumental artifacts in the data (see Appendix \ref{sec:instrumental feature}).

\begin{figure}
	\centering
		\includegraphics[width=0.45 \textwidth]{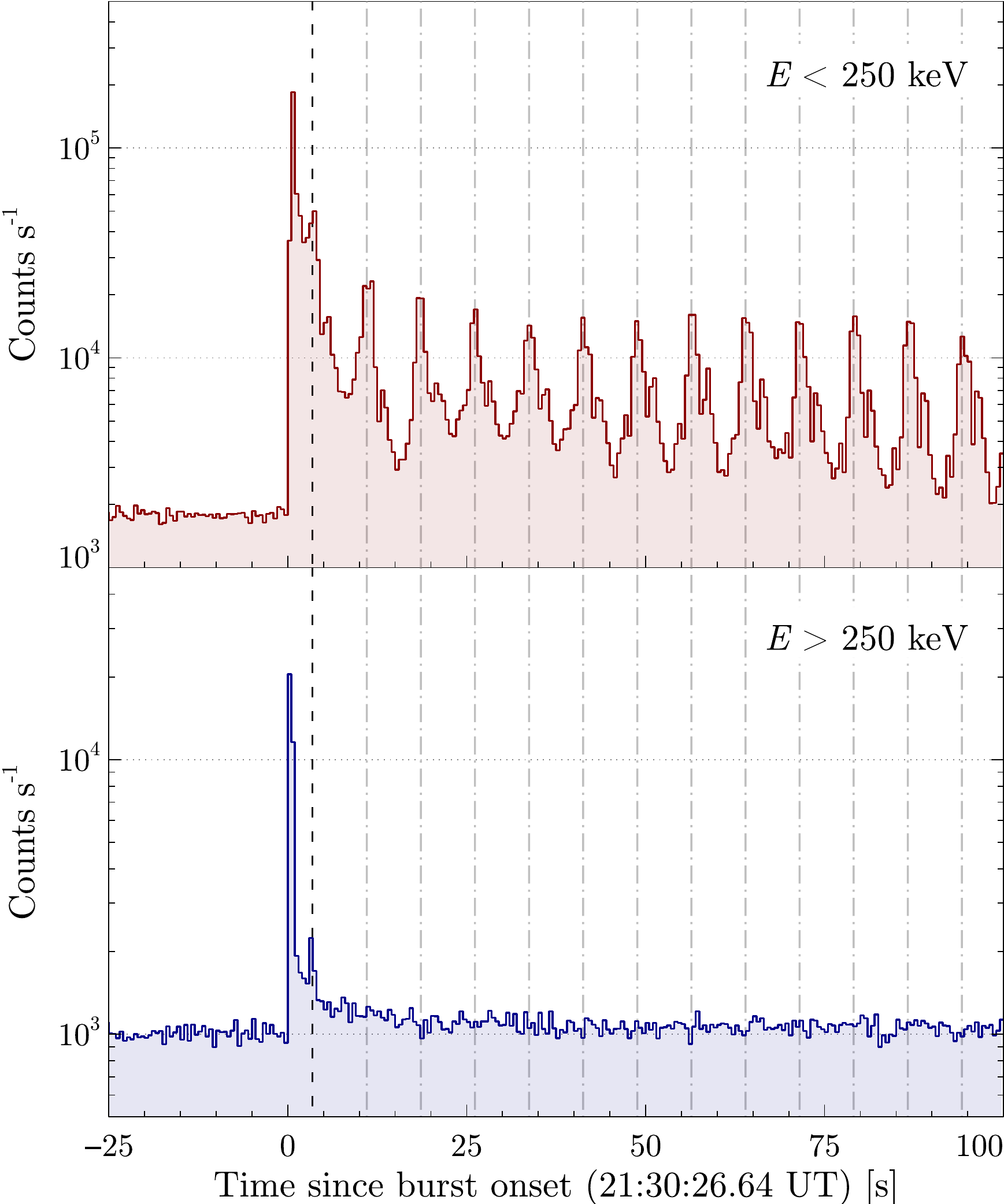}
	\caption{Count rate versus time of the \emph{RHESSI} detectors. The giant flare starts at $t_0=$ 21:30:26.64 UT. The upper (lower) plot displays the low (high) energy count rate, i.e. counts with a recorded energy $<250$ keV ($>250$ keV). The initial hard spike lasts $\sim0.2$ s and is followed by a soft tail with a series of superimposed  pulsations at the spin period of the neutron star, i.e. $P_{\rm NS}\simeq7.56$ s \citep{Woods2007}. The first pulse occurs at $t-t_0\sim3.47$ s, denoted by the vertical dashed line and coincides with an instrumental artifact -- see Appendix \ref{sec:instrumental feature}. The subsequent pulse maxima are denoted by the vertical dash-dotted lines. The pulsations are not apparent in the high energy emission. The plot is truncated at $t-t_0=100$ s. Recorded counts of all detector segments (front and rear) where used in these plots.}\label{fig:lc_1}
\end{figure}

On 2004 December 27 at $t_0=21$:30:26.64 UT, \emph{RHESSI} recorded a GF a mere $\sim5^\circ$ off-axis, that originated from the magnetar SGR 1806--20 (see Fig.~\ref{fig:lc_1}); the third and most energetic magnetar GF ever observed, with total energy (assuming isotropic emission) $\sim10^{46}$ erg \citep{Hurley2005}. A detailed analysis of the obtained \emph{RHESSI} data was presented by \citet{Boggs2007}. In particular, they demonstrated the presence of a nonthermal high-energy emission component ($> 400$ keV) following the onset of the flare that lasts a few tens of seconds. In this section we first reproduce their results regarding the high energy emission, and then further investigate the temporal and spectral properties of the event. 

\subsection{Temporal and spectral properties}\label{sec:temporal and spectral properties}

\begin{figure}
\includegraphics[width=0.45 \textwidth]{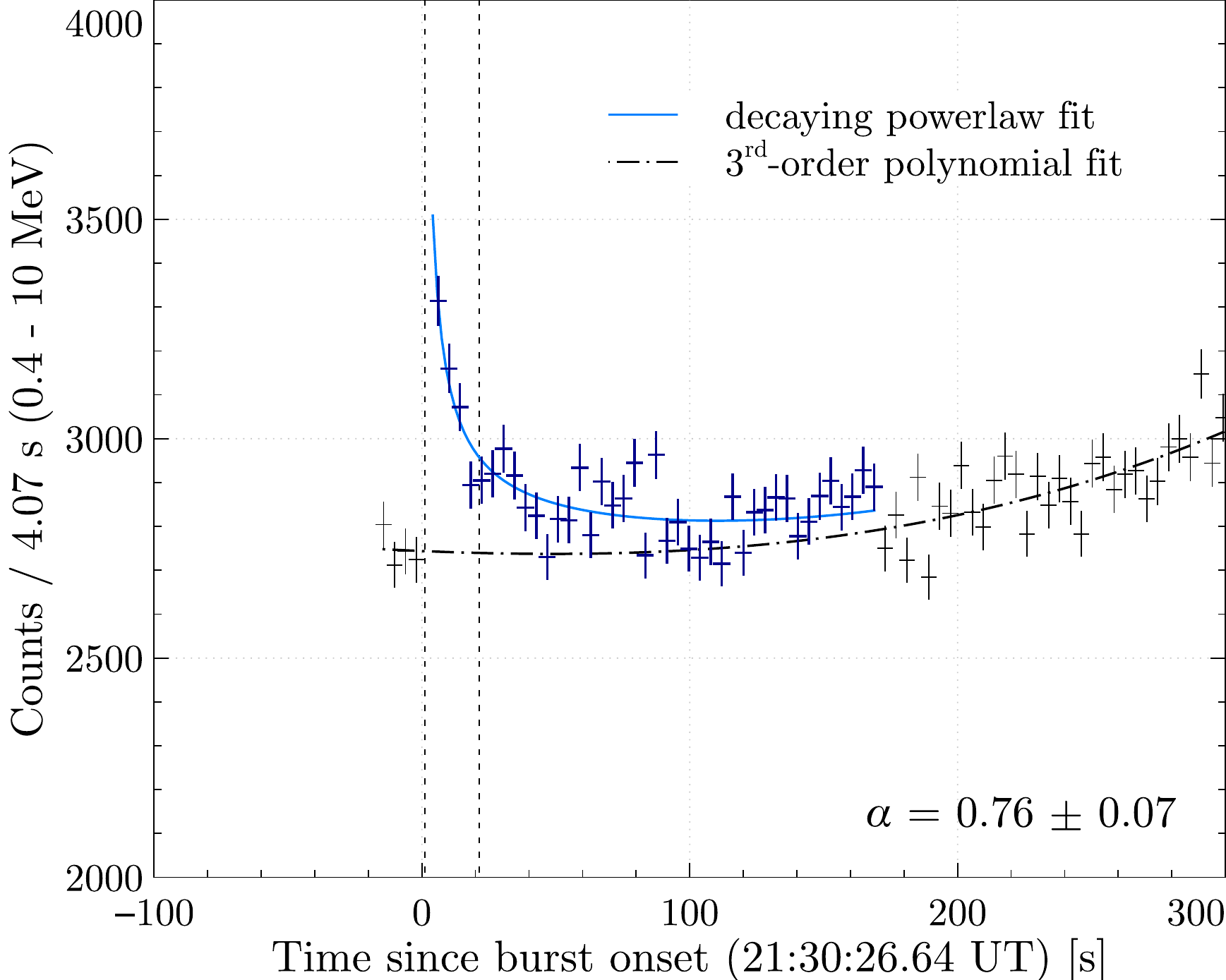}
	\caption{High-energy (0.4--10 MeV) counts (in the rear segments) against time since flare onset $t_0$ in time bins equal to the spacecraft rotation period, i.e. 4.07 s. A high energy excess is present (thick crosses) for $\sim 100$ s above the background (thin crosses). A $3^{\rm{rd}}$-order polynomial was fit to the background counts (dash-dotted curve) and a decaying powerlaw $f(t)\propto t^{-\alpha}$ was fitted to the excess counts, i.e. the background-subtracted counts, with best-fit parameter $\alpha=0.76\pm0.07$ ($\chi^2_{\rm{red}}=1.49,$ 39 dof); the solid curve denotes the fit to the excess counts on top of the fit to the background. The spectrum has been generated from the time interval $t-t_0=1.07-21.42$ s, denoted by the vertical dashed lines. This figure should be compared to Fig. 9 of \citet{Boggs2007}.} \label{fig:lc_2}
\end{figure}

We begin by repeating the analysis performed by \citet{Boggs2007} to confirm the general properties  of the high energy component. The presence of the high-energy excess emission ($0.4-10$ MeV) is demonstrated in Fig.~\ref{fig:lc_2}, where the counts, recorded in the rear segments only, are plotted against time in 4.07 s time bins. Thick crosses denote the high-energy emission associated with the GF starting at $t_0$ and lasting $\sim100$ s. The background emission is represented by thin crosses and fitted with an appropriate smooth function (here a $3^{\rm{rd}}$-order polynomial), i.e. the dash-dotted curve. Consequently, the excess counts were fitted with a decaying powerlaw $f(t)\propto t^{-\alpha}$ shown as the solid curve, with best-fit parameter $\alpha=0.76\pm0.07$ ($\chi^2_{\rm{red}}=1.49,$ 39 dof). This estimate is consistent with the one found by \citet{Boggs2007}, i.e. $\alpha_{\rm B07}=0.68\pm0.04$ ($\chi^2_{\rm{red}}=1.63,$ 39 dof). Fig.~\ref{fig:lc_2} should be compared to Figure 9 in \citet{Boggs2007}.

\begin{figure}
	\includegraphics[width= 0.45 \textwidth]{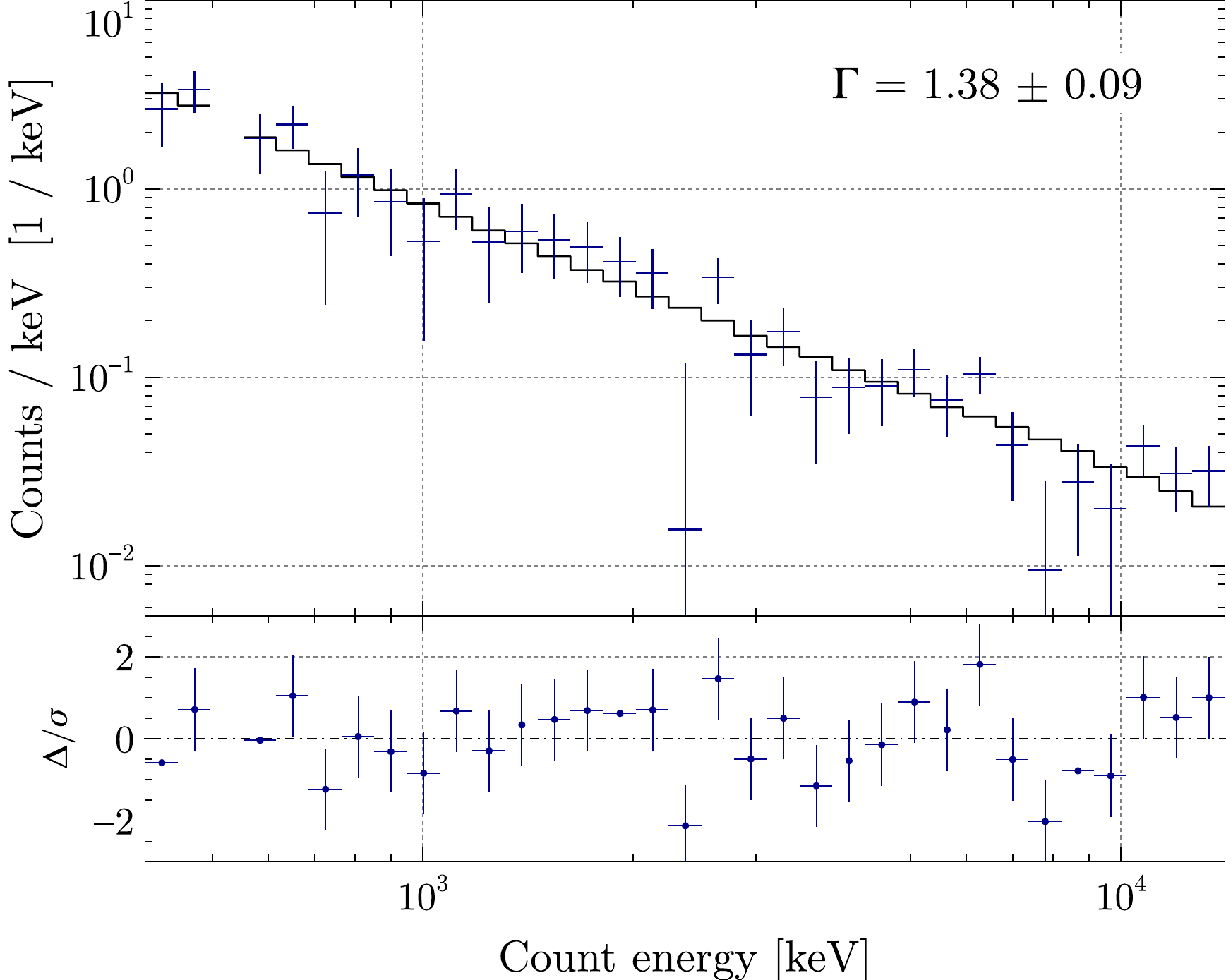}
	\caption{\emph{RHESSI} count energy spectrum integrated over time interval $t-t_0=1.07-21.42$ s. The top panel shows the background subtracted count distribution (crosses) and folded best-fit source photon spectral model [step curve; $\Gamma=1.38\pm0.09$ ($\chi^2_{\rm{red}}= 0.91$, 30 dof)] versus energy. The bottom panel shows the residuals. The energy bin including the 0.511 MeV line was omitted in determining the best-fit photon spectral model.}
\label{fig:spec}
\end{figure}

A high-energy ($0.4-15$ MeV) background subtracted count energy spectrum, shown in Fig.~\ref{fig:spec}, was generated from the integrated time interval\footnote{Our time interval coincides exactly with the one chosen by \citet{Boggs2007}. Note however that they round $t_0$ to the nearest whole second, i.e $t_0^{\rm B07}=$ 21:30:26 UT.}  $t-t_0=1.07-21.42$ s (with $\Delta t_{\rm spec}\equiv20.35$), denoted by the vertical dashed lines in Fig.~\ref{fig:lc_2}. The top panel displays the count energy distribution $dN/dE$ (crosses) of the counts recorded in the detectors rear segments and a folded best-fit source photon spectral model (step curve). The latter was determined via a `forward-fitting' procedure in \texttt{XSPEC} employing the appropriate instrumental response matrix for \emph{RHESSI} (see Fig.~\ref{fig:RHESSI_rm}) observing a source at $5^\circ$ off-axis. The bottom panel shows the residuals of the fit. The bin containing the 511 keV line was excluded from the fitting procedure.

We assumed a simple powerlaw (PL) model for the photon source spectrum $f(\varepsilon)=\mathcal{N}\varepsilon^{-\Gamma}$, where $\varepsilon$ is the photon energy, and obtained the following best-fit parameters: A normalization $\mathcal{N}=4.58^{+5.07}_{-2.52}$ photons keV$^{-1}$ cm$^{-2}$ s$^{-1}$ at 1 keV and photon index $\Gamma=1.38\pm0.09$, with fit statistic $\chi^2_{\rm{red}}= 0.91$, 30 dof. Accordingly, we may infer a photon flux integrated over the energy range $0.4-15$ MeV of $F=(4.2\pm0.3)\times10^{-6}$ erg cm$^{-2}$ s$^{-1}$, with time integrated fluence over $\Delta t_{\rm spec}$ of $\mathcal{F}=(8.6\pm0.6)\times10^{-5}$ erg cm$^{-2}$. This result deviates from the fluence found by \citet{Boggs2007}, i.e $\mathcal{F}=(9.8\pm0.1)\times10^{-5}$ erg cm$^{-2}$. We have not been able to fully determine the cause of this slight difference, since some relevant details of the data analysis done for the 2007 publication are not given. After consultation with the authors, however, we believe that the differences likely arise from the use of a distinct background extraction procedure resulting in a different background-subtracted source spectrum and adopting a slightly more up to date instrument response matrix that may modify the best-fit parameters of the photon spectrum model and consequently the inferred photon fluence (E. Bellm, private communication).

Having confirmed the results of \citet{Boggs2007} regarding in particular the existence and general properties of the nonthermal hard component, we will now continue to investigate the pulsed emission. In particular we would like to know how pulsed is the high-energy emission.

\subsection{Pulsed emission}\label{sec:pulsed emission}

From Fig. \ref{fig:lc_1} we notice that a strong pulsed fraction seems to be absent in the high energy emission, as compared to the low energy emission \citep{Boggs2007}. The degree to which the emission is pulsed in all of the bands will be important when we come to consider mechanisms that might generate the high energy component. In this section we will examine whether the apparent drop off in pulsed amplitude at the highest energies is genuine, or whether it could be an artifact of lower signal to noise in the higher energy band.  

\subsubsection{Issues with standard periodicity detection approaches}
\begin{figure}
	\includegraphics[width= 0.45 \textwidth]{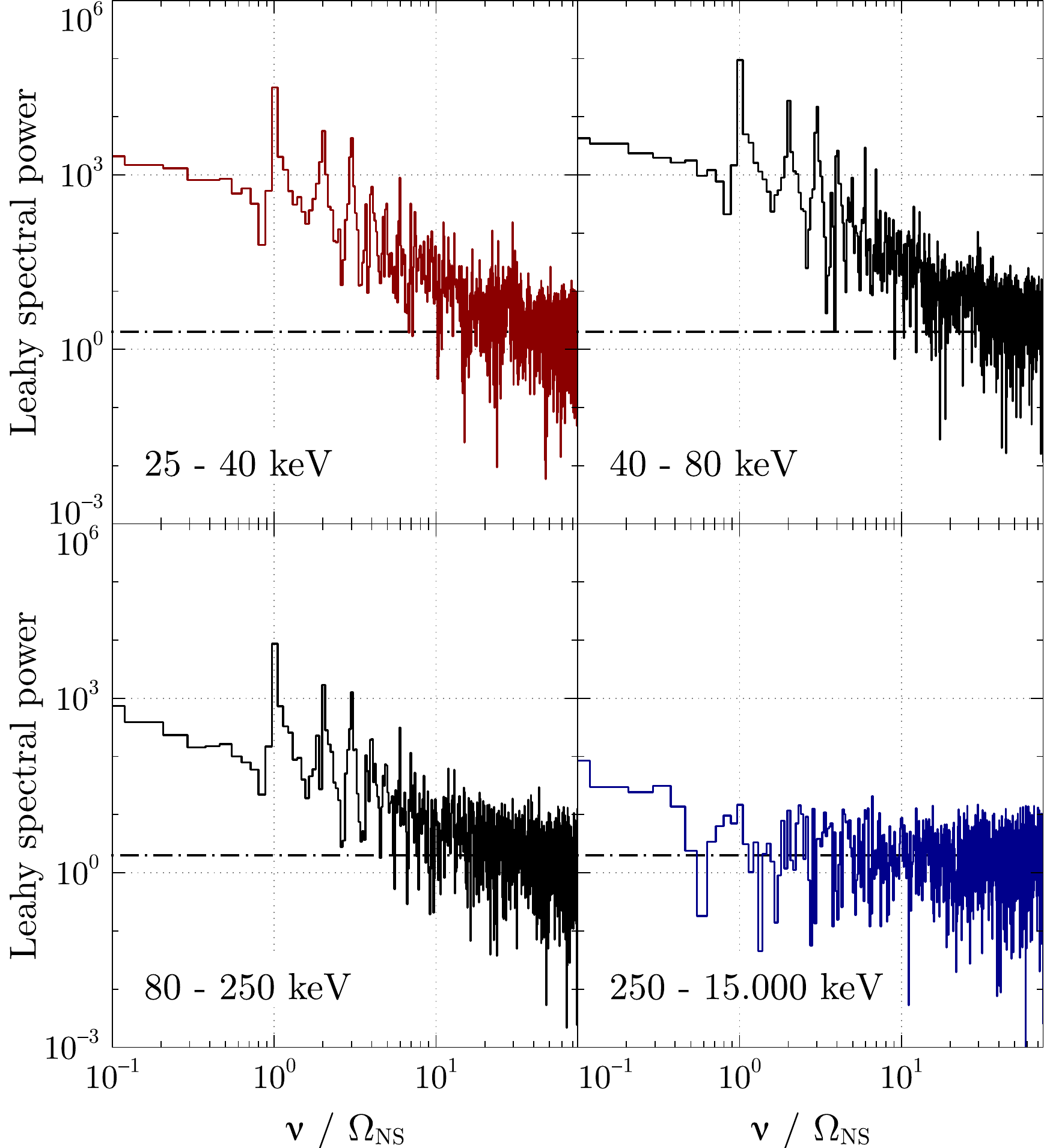}
	\caption{Power spectral density plots of the separate energy bins ($25-40$ keV, $40-80$ keV, $80-250$ keV, and $250-15,000$ keV). The dash-dotted line denotes the Poisson noise level at $P=2$. Notice the presence of a strong red-noise component in three low energy bins.  The background dominates the highest energy bin, Fig.~\ref{fig:lc_2}, so the red noise is much less pronounced. Furthermore, we can clearly distinguish the fundamental frequency at $\nu/\Omega_{\rm NS}$ in the low energy bins ($< 250$ keV). The high-energy bin ($> 250$ keV) is dominated by noise, making it intricate to detect or negate the presence of any periodic variability.}
\label{fig:psds}
\end{figure}

Comparing the pulsed fraction in low and high energy bands is not straightforward: whether the pulsed fraction is detected (or, indeed, detectable) depends on the details of the detector efficiency as a function of energy, the source energy spectrum and the sky background in the detector as a function of energy.  Here we employ a simple empirical model to test whether the pulsed fraction is constant as a function of energy.  Specifically, we extract light curves in four energy bands ($25-40$ keV, $40-80$ keV, $80-250$ keV, and $250-15,000$ keV) at a time resolution of $50\,\mathrm{ms}$. We consider a segment between $t_0 + 4.2\,\mathrm{s}$ and $t_0 + 94.2\,\mathrm{s}$. The start time of this segment is set roughly by the end time of the first peak of the pulsating phase, since an instrumental artifact coincides with the first pulse maximum (see Fig.~\ref{fig:lc_1}  and Appendix \ref{sec:instrumental feature}). The end time is set approximately by the end of the detectable emission in the highest energy band. The complex time variability of the burst, including an overall decay, a re-brightening and the changing shape of the pulse profile, lead to an equally complex power spectrum including a strong red noise component at low frequencies (see Fig.~\ref{fig:psds}). Consequently, standard periodicity detection algorithms, which rest on the assumption of a white noise background, cannot be used \citep[see e.g.][]{Vaughan2010, Huppenkothen2013}.

\subsubsection{Building Template Models for the Observations}
In order to model the observed emission in all four energy bands, we build a template pulse profile using the brightest band ($40-80$ keV). We first re-bin this light curve to a coarser resolution of $0.1\,\mathrm{s}$ and then fit the binned light curve with a cubic spline function, which allows us to interpolate a smooth version back to the original time resolution. We then extract an estimate of the total background by considering an interval of $26\,\mathrm{s}$ recorded before the burst in the same energy range and  compute the count rate in that interval, assuming that all of the emission recorded in that interval is due to sky background. Under the additional assumption that the sky background is constant, we  subtract off the counts per bin due to background from the template. By dividing this smooth version by the total number of source photons (i.e. the total number of photons observed in this band, minus the total number of photons assumed to be due to background) in this energy band, we arrive at a template to use for further analysis.

In order to compare this template to the observations, we use the spectrum of the flare interval as well as the pre-flare interval to extract the integrated photon flux in the detector for source and background, respectively. The total photon flux from the flare interval includes contributions from both source and sky background. In order to get the source flux only, we subtract the time scaled total number of photons expected from the sky background based on the pre-flare interval in the same energy band. We then construct a template light curve by multiplying the pulse template by the total source photons in each bin, and adding a constant sky background that integrates to the total sky background as extracted from the spectrum.  This procedure results in a model light curve for each bin, appropriately scaled by the number of photons recorded in the detector, thus it takes the effects of the detector efficiency and the source energy spectrum into account.

\subsubsection{Bayesian Model Comparison}
The ultimate goal is to test whether we can confidently exclude a model where the pulsed fraction is constant with energy. Note that this is conceptually different from trying to  to detect whether there are any pulsations present in the data. In the latter case, we would presume no other knowledge of the problem and simply ask the question whether the data supports the hypothesis of a (quasi-)periodicity in a given energy bin, without taking the other energy bins into account. Here, however, we are more interested in whether the data are consistent with a model where a periodicity is present in all four energy bins, but modulated by the changes in source spectrum and sky background as a function of energy, so that it might still go undetected with other period detection algorithms. To do so, we introduce an amplitude parameter $A_i$ for each energy bin $E_i$, which scales the pulse profile. This allows us to set up two models encoding the hypotheses we are interested in. In model 1 ($M_1$), we assume the pulsed fraction is constant with energy, and any decrease in observed pulse fraction is caused by the energy dependence of source spectrum and sky background. Thus, in $M_1$ we define that the amplitudes $A_i$ equal a single constant parameter $C$ for all $i$, i.e. $A_i = C \, \forall i$. In model 2 ($M_2$), we allow the pulse amplitude to vary as a function of energy. Because we have no \textit{a priori} knowledge of the functional form of this dependence, we allow the amplitude to vary independently in each energy bin. We parametrize the amplitude as $A_i = C + a_i \, \forall i$, which gives an additional set of $N$ parameters $\{a_i\}_{i=1}^{N}$ for  $N=4$ light curves.  While we could have parametrized $A_i$ directly and reduced a parameter, this parametrization allows for the models to be nested (where the simpler model is a special case of the more complex model with $a_i = 0 \, \forall i$), which is convenient for model comparison purposes.

Using these two parametrizations, we build two models for the observations in a Bayesian framework and use Markov Chain Monte Carlo (MCMC) to sample their parameters. Because the pulse shape may be mildly energy dependent, we do not compare the light curve template directly to the observed light curves, but instead produce power density spectra (PDS) of the observed light curves as well as the model light curves.  We add a power law component (with normalization $A_{\mathrm{PL}, i}$ and index $\Gamma_i$) to each transformed model PDS to account for any red noise not modeled by the pulse shape itself as well as a constant $w_i$ to account for Poisson statistics at high frequencies. This adds a  set of twelve parameters $\{\Gamma_i, A_{\mathrm{PL}, i}, w_i\}_{i=1}^{N}$ to each model.

Because the GF light curves were extracted in energy bands that do not overlap, we can consider them to be statistically independent, and can thus define a likelihood over all four light curves as the product of the individual likelihoods for each energy band.  This, along with reasonable priors on the parameters (see Appendix \ref{sec:pulsemodel} for details of the model and procedure), allows us to sample from the full model. 

\begin{figure}
	\includegraphics[width= 0.45 \textwidth]{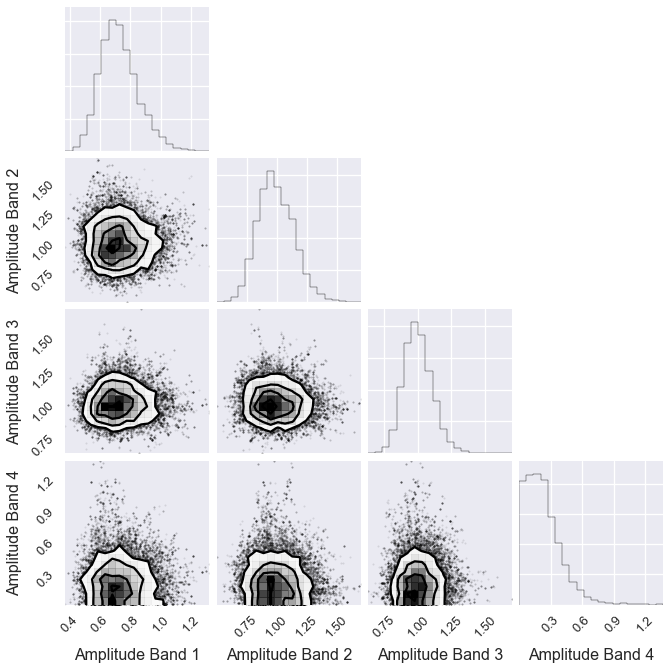}
	\caption{Posterior distributions for the pulse amplitudes in the four different bands. The distributions were calculated as $A_i = C + a_i$ for each band $i$ and each sample in the MCMC chain.}
\label{fig:pulse_posteriors}
\end{figure}

\subsubsection{Results}
We use MCMC in the form of \textit{emcee} \citep{Foremanmackey2013} to sample the posterior distributions of both $M_1$ and $M_2$ and find that the posterior distributions for $M_1$ and $M_2$ are well-constrained.  In order to compare $M_1$ and $M_2$ more formally, we exploit the nested nature of the models and use the Savage-Dickey Density (SDDR; defined in Appendix \ref{sec:pulsemodel}) to approximate the ratio of marginalized posterior probability densities of $M_1$ to $M_2$ (i.e.\ the Bayes factor).
Using the SDDR, we find a logarithmic Bayes factor of $-1.6$, corresponding to a $74\%$ probability of $M_2$ being true. Following Jeffrey's scale \citep{Jeffreys1998} and assuming equal prior probabilities for $M_1$ and $M_2$ (i.e.\ $p(M_1) = p(M_2) = 0.5$), we interpret this as very weak evidence in favor of the model with variable pulse amplitude. In Fig.~\ref{fig:pulse_posteriors}, we present the posterior distributions for the pulse amplitudes. We note that while the posterior distributions for the pulse amplitudes are fairly wide, there is some indication that the pulse amplitude might not be constant here, too. In particular, the pulse amplitude for the highest-energy bin is consistent with this emission not being pulsed. Furthermore, the emission in the lowest energy bin also appears to be less pulsed than the two middle bins.

We note, however, that this result depends quite strongly on prior choice: the Bayes factor trades off information content in the data with the volume of parameter space allowed by the prior. If the latter is very large for the additional parameters in $M_2$, the automatic penalty imposed may lead to a preference of the simpler model if the data are not exceptionally informative. Replacing the uniform prior for $a_i$ with a Gaussian prior centered on zero with a standard deviation of $0.2$ changes our conclusions quite drastically: in this case, there is moderate evidence for the model with constant pulse amplitude $M_1$ (with posterior odds of $0.94$). Given that even a fairly wide Gaussian prior can change the model's inference this drastically, we conclude that the data are not very informative. If it were, the likelihood would dampen the sensitivity of our conclusions to the choice of prior, and the model would converge to the same conclusions. This is also reflected in the likelihood ratio which only weakly prefers the model where the pulse amplitude depends on energy.

\section{Global flare model: A highly extended corona}
\label{sec:icmodel}

\begin{figure*}
	\centering
		\includegraphics[width= \textwidth]{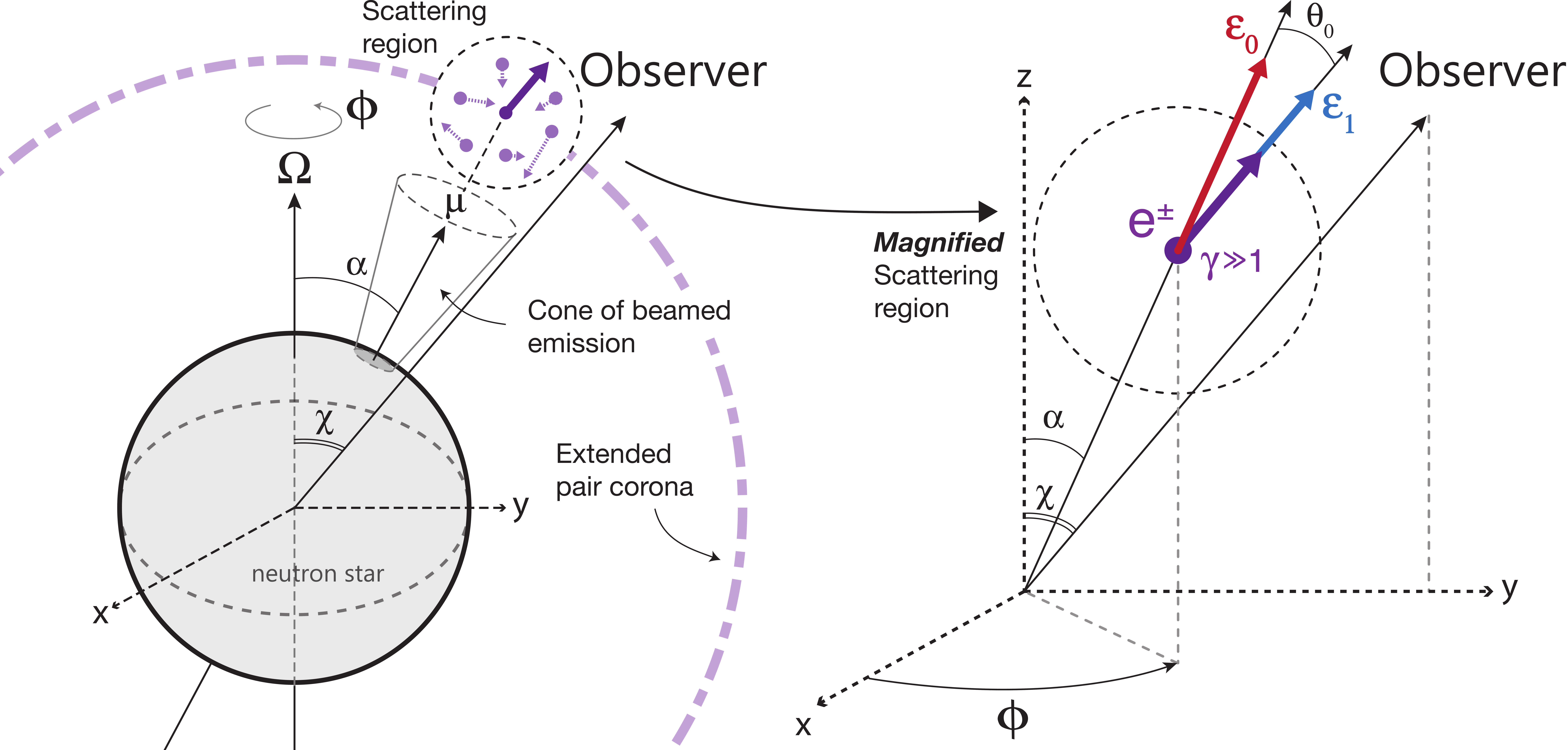}
	\caption{Schematic of the inverse Compton scattering configuration (not to scale). During the pulsation phase of the giant flare a cone of beamed emission emerging from the evaporating fireball sweeps through the line of sight of the observer. Moreover, a portion of this anisotropic low energy (i.e. $< 250$ keV) emission proceeds to upscatter in a region that is part of an extended corona of isotropic relativistic pairs that envelops the NS, resulting in the high energy spectral component ($> 400$ keV). Here, $\boldsymbol{\varepsilon_0}$ and $\boldsymbol{\varepsilon_1}$ respectively denote the incident photon unit vector (red thick arrow) and unit vector of the upscattered photon in the direction of the observer (blue thick arrow), $\chi$ is the angle between the rotation vector of the neutron star and the observer, $\cos\alpha=\boldsymbol{\mu}\cdot\boldsymbol{\Omega}$ denotes the angle between the rotation and magnetic dipole vector of the NS (or indeed the normal vector at the base of the fireball), $\theta_0$ represents the incident photon angle, i.e. the angle between $\boldsymbol{\varepsilon_0}$ and the pair's direction of motion (in turn given by the purple thick arrow). Due to the rotation of the neutron star, the orientation of $\boldsymbol{\mu}(\phi)$ will vary periodically with respect to the observer, such that $\theta_0(\phi)$ will depend on the phase. From simulating inverse Compton spectra we find that the photon spectral index $\Gamma$ depends strongly on $\theta_0(\phi)$  (see Fig.~\ref{fig:IC_spectra}) and will therefore also be phase-dependent.}
\label{fig:config}
\end{figure*}

In this section we explore the model proposed by \citet{Boggs2007} to explain the high energy nonthermal component during the peak decay. They hypothesize that this component originates in a highly extended corona, that has been driven by the hyper-Eddington luminosities, where the synchrotron cooling of the scattering particles is inefficient. The seed photons that are upscattered come from the jets of radiation that emanate from the base of the trapped pair-plasma fireball. They argue further that the lack of apparent pulsations in the high energy light curve is consistent with this picture. Our analysis of the afterglow emission indeed shows that a model where the pulsed fraction decreases for the high energy emission is preferred.

\subsection{Inverse-Compton Scattering}

Here we wish to investigate whether the model proposed by \citet{Feroci2001} and \citet{Boggs2007} can reproduce the observed high energy spectral component and furthermore whether such a process would isotropise or increase the degree of beaming of the beamed incident radiation emerging from the base of the fireball.  We will make use of inverse Compton (IC) scattering calculations involving an anisotropic incident photon beam undertaken by \citet{Dubus2008} and Cerutti (2010) in investigating the phase-dependent emission from $\gamma$-ray binaries. Here however we consider the following scenario, illustrated in Fig. \ref{fig:config}. During the pulsation phase of the GF light curve, we consider the configuration wherein the pulsed soft X-ray emission is caused by a cone of beamed emission aligned with the magnetic dipole vector $\boldsymbol{\mu}$ of the NS\footnote{The base of the fireball may not be aligned with the dipole vector of the NS. In that case $\boldsymbol{\mu}$ represents the normal vector to the NS surface at the location of the base of the fireball.} that moves in and out of the line of sight (parameterised by  $\chi$, i.e. the angle between the observer and rotational vector $\boldsymbol{\Omega}$ of the NS) as the underlying NS rotates. We subsequently assess the model where a portion of this beamed emission irradiates a \emph{highly} extended corona comprising an isotropic distribution of pairs that envelops the NS. These pairs in turn upscatter, through inverse Compton scattering, the incident low energy photons to higher energies ($>$ tens of MeV). 

We assume that the pairs are relativistic ($\gamma\gtrsim10$), such that the outgoing photons will be beamed primarily, over an angle $\sim1/\gamma$, in the direction of motion of the charged particles. Accordingly, we only observe the Comptonized emission scattering from charged particles that propagate in the direction of the observer. Consequently, we solely observe photons that have interacted with a charged particle at $\theta_0$, which is given by 
\begin{equation}\label{eq:theta_0}
\cos\theta_0(\phi)=\cos\alpha\cos\chi+\sin\alpha\sin\chi\cos\phi\equiv\mu_0(\phi),
\end{equation}
where $\cos\alpha=\boldsymbol{\mu}\cdot\boldsymbol{\Omega}$ denotes the angle between the rotation and magnetic dipole vector of the NS, and $\phi$ represents the rotational phase of the NS. 

In this configuration, the extended pair corona is hypothesized to be uniform. We will test this conjecture and discuss its validity in sections \ref{sec:IC cooling time and size of the corona} and \ref{sec:Synchrotron cooling time and lower limit to the size of the corona}, focusing mainly on its size constraints. The extended corona is illuminated at different longitudes, i.e. the scattering regions, as the soft X-ray beam sweeps round with $\phi$. Accordingly, we approximate the incident photon spectrum as a blackbody distribution of constant temperature $k_{\rm B}T_{\rm bb}$. Furthermore, we assume that it is highly anisotropic, i.e. we model the emission cone as a entirely radially outward directed beam irradiating a scattering region from below at each instant in time. The seed photon distribution is therefore given by
\begin{equation}\label{eq:seed photon energy distribution}
\frac{dn_{\rm X}}{d\varepsilon_0}=\frac{2}{h^3c^3}\frac{\varepsilon_0^2}{\exp\left[\frac{\varepsilon_0}{k_{\rm B}T_{\rm bb}}\right]-1},
\end{equation}
where $n_{\rm X}$ is the soft X-ray photon number density, $\varepsilon_0$ is the incident photon energy, $h$ is Planck's constant, and $c$ is the speed of light. A fraction of these photons interacts kinematically with the relativistic pairs in the extended corona. We assume a simple isotropic power law energy distribution for the pairs up to a maximum energy\footnote{For now, we have no physical reason to set a lower limit to the energy of the pairs. The distribution extends down to $\gamma_- = 1$. Our Klein-Nishina kernel [Eq.~(\ref{eq: K-N kernel})] and the assumption that we only observe upscattered photons from relativistic pairs that move in the direction of the observer, however both depend on an approximation that requires $\gamma\gg1$. We address this assumption in Section~\ref{sec:Validity of the Klein-Nishina kernel}.} $\gamma_+$,
\begin{equation}\label{eq:isotropic energy distribution of the pairs}
\frac{dN_\pm}{d\gamma}=\frac{1}{4\pi}K_\pm \gamma^{-p}~~~~ \gamma < \gamma_+,
\end{equation}
where $p$ denotes the spectral index of the pair energy distribution and $K_\pm$ represents the normalization in number of pairs per cm$^2$. 

The differential cross section of Compton scattering described by QED is given in the particle rest frame (PRF) by the Klein-Nishina (K-N) equation, 
\begin{equation}\label{eq:K-N cross section}
\frac{d\sigma}{d\Omega_1'd\varepsilon_1'}=\frac{r_e^2}{2}\left(\frac{\varepsilon_1'}{\varepsilon_0'}\right)^2\left(\frac{\varepsilon_1'}{\varepsilon_0'}+\frac{\varepsilon_0'}{\varepsilon_1'}-\sin^2\Theta'\right)\delta\left(\varepsilon_1'-\varepsilon_{1*} '\right),
\end{equation}
where the primed quantities are defined in the PRF, $r_e\equiv e^2/(m_e c^2)$ is the classical electron radius, $\varepsilon_1$ is the energy of the outgoing photon, $\Theta$ is the angle between the incident and outgoing photon, $\delta$ denotes the Dirac distribution, and the Compton formula is given by
\begin{equation}
\varepsilon_{1*}'=\frac{\varepsilon_0'}{1+\frac{\varepsilon_0'}{m_\pm c^2}\left(1-\mu_{\Theta}'\right)},
\end{equation}
where $\mu_\Theta'=\cos\Theta'=\mu_1'\mu_0'+(1-\mu_1')^{1/2}(1-\mu_0')^{1/2}\cos(\phi_1'-\phi_0')$ and $\mu_i=\cos\theta_i$, with $i\in\{0,1\}$. We may translate the photon energies between the PRF and the lab frame through the relativistic Doppler shift equations:
\begin{align}
&\varepsilon_0'=\gamma\left(1-\beta\mu_0\right)\varepsilon_0,\\
&\varepsilon_1'=\gamma\left(1-\beta\mu_1\right)\varepsilon_1,
\end{align}
where $\beta = v/c = (1-\gamma^{-2})^{1/2}$ is particle's velocity in units of $c$. Furthermore, the angles $\mu_i$, with $i\in\{0,1\}$, translate according to the formula for the aberration of light, 
\begin{equation}
\mu_i'=\frac{\mu_i-\beta}{1-\beta\mu_i}.
\end{equation} 
The \emph{normalised} anisotropic seed photon density for a monochromatic beam is given by
\begin{equation}
\frac{dn}{d\varepsilon d\Omega}=\delta\left(\varepsilon-\varepsilon_0\right)\delta\left(\mu-\mu_0\right)\delta\left(\phi-\phi_0\right),
\end{equation}
which in the PRF becomes 
\begin{equation}\label{eq:anisotropic seed photon density in PRF}
\frac{dn'}{d\varepsilon' d\Omega'}=\gamma(1-\beta\mu)\delta\left(\varepsilon'-\varepsilon_0'\right)\delta\left(\mu'-\mu_0'\right)\delta\left(\phi'-\phi_0'\right).
\end{equation}
In the lab frame, the anisotropic K-N kernel is then given by the following relation
\begin{equation}\label{eq:  K-N kernel integral form}
\frac{dN}{dt d\varepsilon_1 d\Omega_1}=\frac{c}{\gamma^2(1-\beta\mu_1)}\iint\frac{dn'}{d\varepsilon' d\Omega'}\frac{d\sigma}{d\Omega_1'd\varepsilon_1'}d\varepsilon' d\Omega'.
\end{equation}
Inserting equations Eq.~(\ref{eq:K-N cross section}), Eq.~(\ref{eq:anisotropic seed photon density in PRF}), and making use of the following approximation $\mu_\Theta'\approx\mu_1'\mu_0'$ valid for $\gamma\gg1$, one can write the anisotropic kernel as such
\begin{align}\label{eq: K-N kernel}
\frac{dN}{dtd\varepsilon_1}\simeq&\frac{\pi r_e^2 c}{\gamma} \xi\frac{1-\beta\mu_0}{1-\beta \mu_1^*}\left\{1+\left(\frac{\mu_1^*-\beta}{1-\beta \mu_1^*}\mu_0'\right)^2\right.\nonumber\\
&\left.+\left(\frac{\gamma \varepsilon_1}{m_\pm c^2}\frac{1+\beta\mu_0'-(\beta+\mu_0')\mu_1^*}{\left\{1-\frac{\gamma \varepsilon_1}{m_\pm c^2}\left[1+\beta\mu_0'-(\beta+\mu_0')\mu_1^*\right]\right\}^{1/2}}\right)^2\right\},
\end{align}
with 
\begin{equation}
\xi \equiv\frac{\left\{1-\frac{\gamma\varepsilon_1}{m_\pm c^2}\left[1+\beta\mu_0'-\left(\beta+\mu_0'\right)\mu_1^*\right]\right\}^2}{\left|\beta\gamma\varepsilon_1+\frac{\varepsilon^2_1}{m_\pm  c^2}\mu_0'\right|},
\end{equation}
and the outgoing photon angle
\begin{equation}
\mu_1^*\simeq\frac{1-\frac{\varepsilon_0}{\varepsilon_1}(1-\beta\mu_0)+\frac{\varepsilon_0}{\gamma m_\pm  c^2}}{\beta+\frac{\varepsilon_0\mu_0}{\gamma m_\pm  c^2}}.
\end{equation}
To attain the emitted spectrum we need to consider the seed photon energy distribution Eq.~(\ref{eq:seed photon energy distribution}) and isotropic energy distribution of the scattering pairs Eq.~(\ref{eq:isotropic energy distribution of the pairs}) and proceed to numerically evaluate the following expression at each emitted photon energy $\varepsilon_1$,
\begin{equation}\label{eq:emitted photon spectrum integration}
\frac{dN}{dtd\varepsilon_1}=\iint\frac{dN_\pm}{d\gamma}\frac{dn_{\rm X}}{d\varepsilon_0}\frac{dN}{dtd\varepsilon_1}d\varepsilon_0d\gamma.
\end{equation}
The input parameters of the simulation are $T_{\rm bb}$, $K_\pm$, $p$, $\gamma_+$, and $\theta_0(\phi)$ where the latter may vary with the rotational phase $\phi$ of the NS, depending on fixed values for $\alpha$ and $\chi$ -- see Eq.~(\ref{eq:theta_0}). The former parameters however are assumed to be constant, i.e. we assume a homogeneous energy distribution for the relativistic pairs in the extended corona, characterized by $K_\pm$, $p$, and $\gamma_{+}$, and that a given scattering region is always irradiated by the same incident photon spectrum, determined by $T_{\rm bb}$. 

\subsection{Results of the IC model}

\subsubsection{The IC spectra}

We investigated the effect of varying the input parameters within a physically reasonable space has on the emitted spectrum, in the photon energy range $300 - 17,000$ keV. We found that varying $p$ changes the hardness of the emitted spectrum notably, whereas the remaining parameters only alter the normalization or display no apparent change in this energy range. For $\gamma_+\lesssim10^2$ the spectrum cuts off exponentially at roughly $\lesssim20$ MeV.

\begin{figure}
	\centering
		\includegraphics[width=0.45 \textwidth]{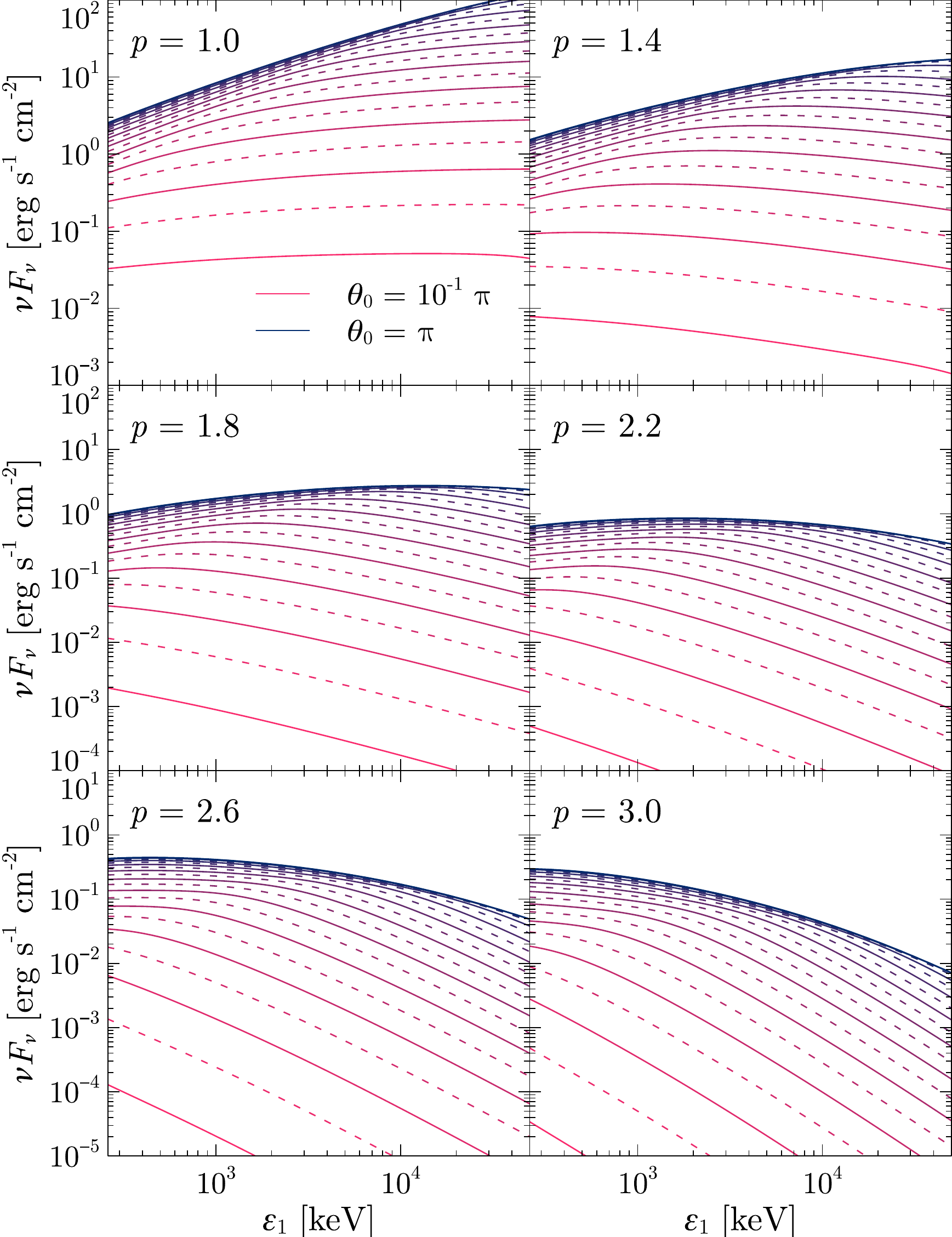}
	\caption{Simulated high-energy inverse Compton emission spectra. The separate graphs show the emission spectra for varying spectral index of the pair energy distribution $p$. The curves per graph represent the emission spectra for different values of the incident photon angle $\theta_0$ ranging from $10^{-1}\pi$ to $\pi$ in steps of $4.5\times10^{-2}\pi$; with $\gamma_+=10^4$ and $k_{\rm B}T_{\rm bb}=10$ keV. Note that the spectra below $\lesssim1$ MeV, and in particular for small values of $\theta_0$, may deviate due to the assumption that $\gamma\gg1$.}\label{fig:IC_spectra}
\end{figure}

Results of the integration of Eq.~(\ref{eq:emitted photon spectrum integration}) are shown in Fig.~\ref{fig:IC_spectra}. Each sub figure exhibits a range of values for $\theta_0=10^{-1}\pi-\pi$ in steps of $4.5\times 10^{-2} \pi$ for a set of values of the spectral index of the pair energy distribution $p\in\{1,~1.4,~1.8,~2.2,~2.6,~3\}$. Note that the slope of the emitted spectra increases as $p$ decreases. We also see that the spectrum is harder with increasing incoming photon angle $\theta_0$, i.e. as the collision with the particle becomes increasingly head-on. 

\begin{figure}
	\includegraphics[width= 0.45 \textwidth]{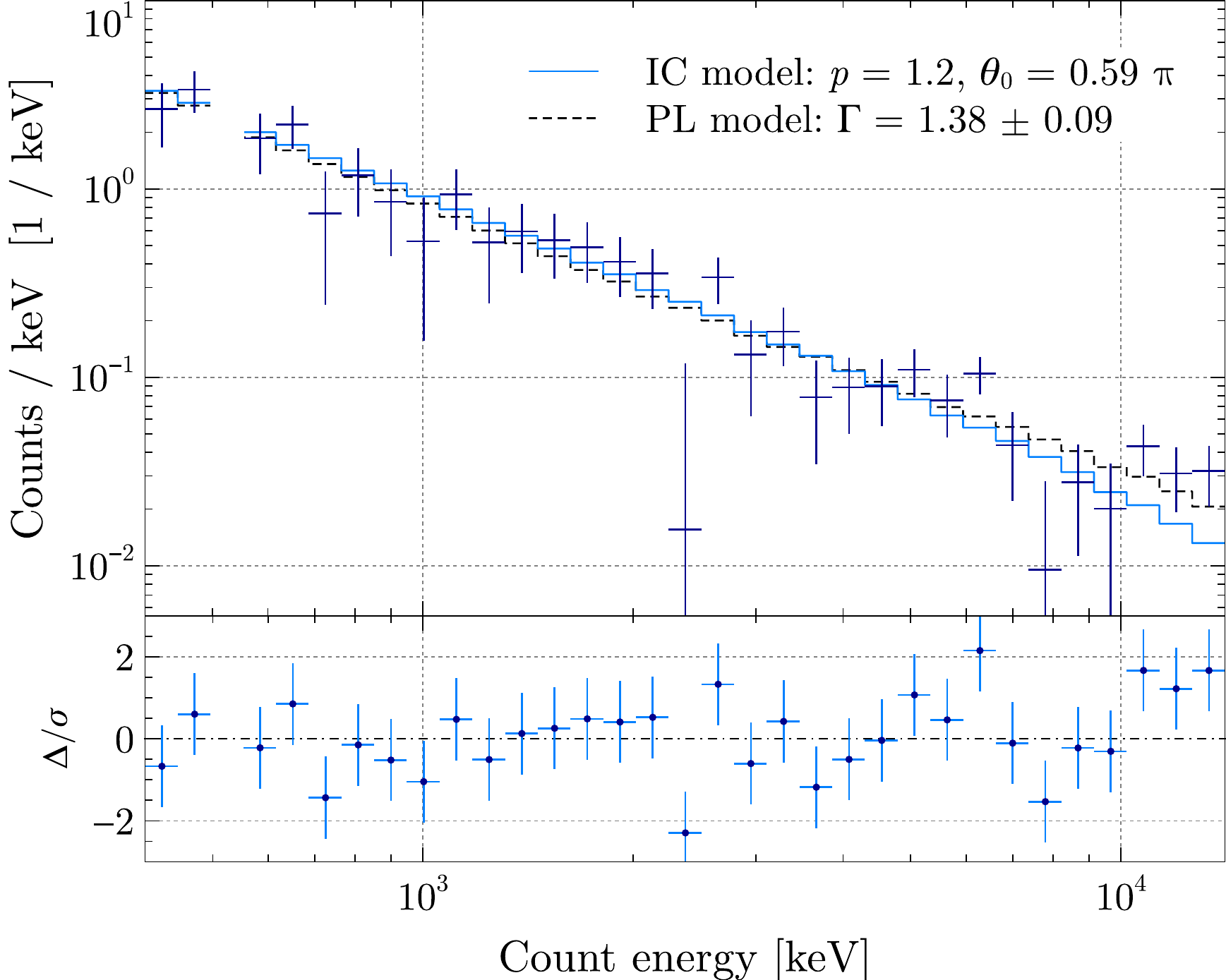}
	\caption{\emph{RHESSI} count energy spectrum integrated over time interval $t-t_0=1.07-21.42$ s. The top panel shows the background subtracted count distribution (crosses), best-fit inverse Compton model [step curve; $p=1.2$ and $\theta_0=0.59\pi$ ($\chi_{\rm red}^{2}=1.00$, 31 dof)], and best-fit powerlaw photon spectral model [dashed step curve; $\Gamma=1.38\pm0.09$ ($\chi^2_{\rm{red}}= 0.91$, 30 dof)] versus energy. The bottom panel shows the residuals of the inverse Compton model. The energy bin including the 0.511 MeV line was omitted in determining the best-fit IC model.}
\label{fig:IC_spec_fit}
\end{figure}

We fit the IC model to the observed data for fixed values of $p$, varying $\theta_0$. Best-fit results are listed in Table~\ref{tab:fit results}. We find that we can attain $\Delta\chi_{\rm red}^2=1.00$ for $p=1-1.4$ and a poor fit for $p=2$ (and $\theta_0=0.81\pi$), with $\Delta\chi_{\rm red}^2=1.90$ and 31 dof.
\begin{table}
\caption{Results of the inverse Compton model fit to the observed \emph{RHESSI} counts for fixed values of $p$ and varying $\theta_0$. Note that the best-fit power law model was described by $\Gamma=1.38\pm0.09$ with $\Delta\chi_{\rm red}^2=0.91$ and 30 dof.}
\centering
\begin{tabular}{ccc}
\hline
$p$  & $\theta_0 / \pi$ & $\Delta\chi_{\rm red}^2$ (31 dof) \\
\hline
\hline
1.0 & 0.46 & 1.00\\
1.1 & 0.53 & 1.00\\
1.2 & 0.59 & 1.00\\
1.3 & 0.65 & 1.00\\
1.4 & 0.73 & 1.00\\
1.5 & 0.83 & 1.06$^{\rm a}$\\
1.6 & 0.83 & 1.18$^{\rm a}$\\
2.0 & 0.81 & 1.90$^{\rm a}$\\
\hline
\end{tabular}\begin{flushleft}{\small $^{\rm a}$Lowest obtainable values for $\Delta\chi^2$ for the respective values of $p$.}\end{flushleft}
\label{tab:fit results}
\end{table}
For comparison we plot the IC spectrum for $p=1.2$ and $\theta_0=0.59\pi$ and the best-fit power law in Fig.~\ref{fig:IC_spec_fit}. If indeed the observed data are produced by Comptonized emission, we find that the energy distribution of the scatterers is required to be very hard to reproduce the observed spectrum, such that the acceleration mechanism must be very efficient $p\lesssim1.5$. For instance, values of  $p = 1-1.2$ may be attained through magnetic reconnection in the ultra-relativistic regime, i.e. where the energy density of the magnetic field dominates over the  particle energy density $\epsilon_B/\epsilon_p\gg1$ \citep{Sironi2011,Guo2014}. These particles may however cool due to strong synchrotron radiation, before being able to Compton upscatter the seed photons (see Section~\ref{sec:Discussion and review of assumptions}). 

If we consider the case where $p=1.2$, $\theta_0=0.59\pi$, and $\gamma_+=10^2$, we find that $K_\pm=(1.03\pm0.07)\times10^{24}$ pairs cm$^{-2}$. Here, we have assumed a typical size for the NS, i.e. $R_{\rm NS}=10^6$ cm and the source to be a distance $d=8.7$ kpc from the observer \citep{Bibby2008}. Subsequently, with Eq.~(\ref{eq:isotropic energy distribution of the pairs}) we can estimate the required integrated energy of the pairs (with $1<\gamma<10^2$) in the scattering region,
\begin{equation}\label{eq:energy in scattering region}
E_\pm=m_\pm c^2\int_{\gamma_-}^{\gamma^+}\gamma\frac{dN_\pm}{d\gamma}d\gamma\simeq3.3\times10^{18}~\text{erg cm}^{-2}.
\end{equation}

\subsubsection{The dependency of the IC spectrum on the phase}

When it comes to studying the pulsed fraction, comparing the simulated spectra to the data of the GF is hindered by the fact that we have no specific information on the angles $\chi$ and $\alpha$, which both determine how $\theta_0$ will vary with $\phi$. What we do know however is that if the low energy emission is strongly beamed and pulsed ($\alpha \notin\{0,\pi\}$), which it indeed should be for SGR 1806--20, then $\theta_0$ will vary strongly as well according to the IC model. The strongest variation in the slope of the emitted spectrum due the variation of $\theta_0(\phi)$ will occur when $\alpha=\chi=\pi/2$.

\begin{figure*}
	\centering
		\includegraphics[width=\textwidth]{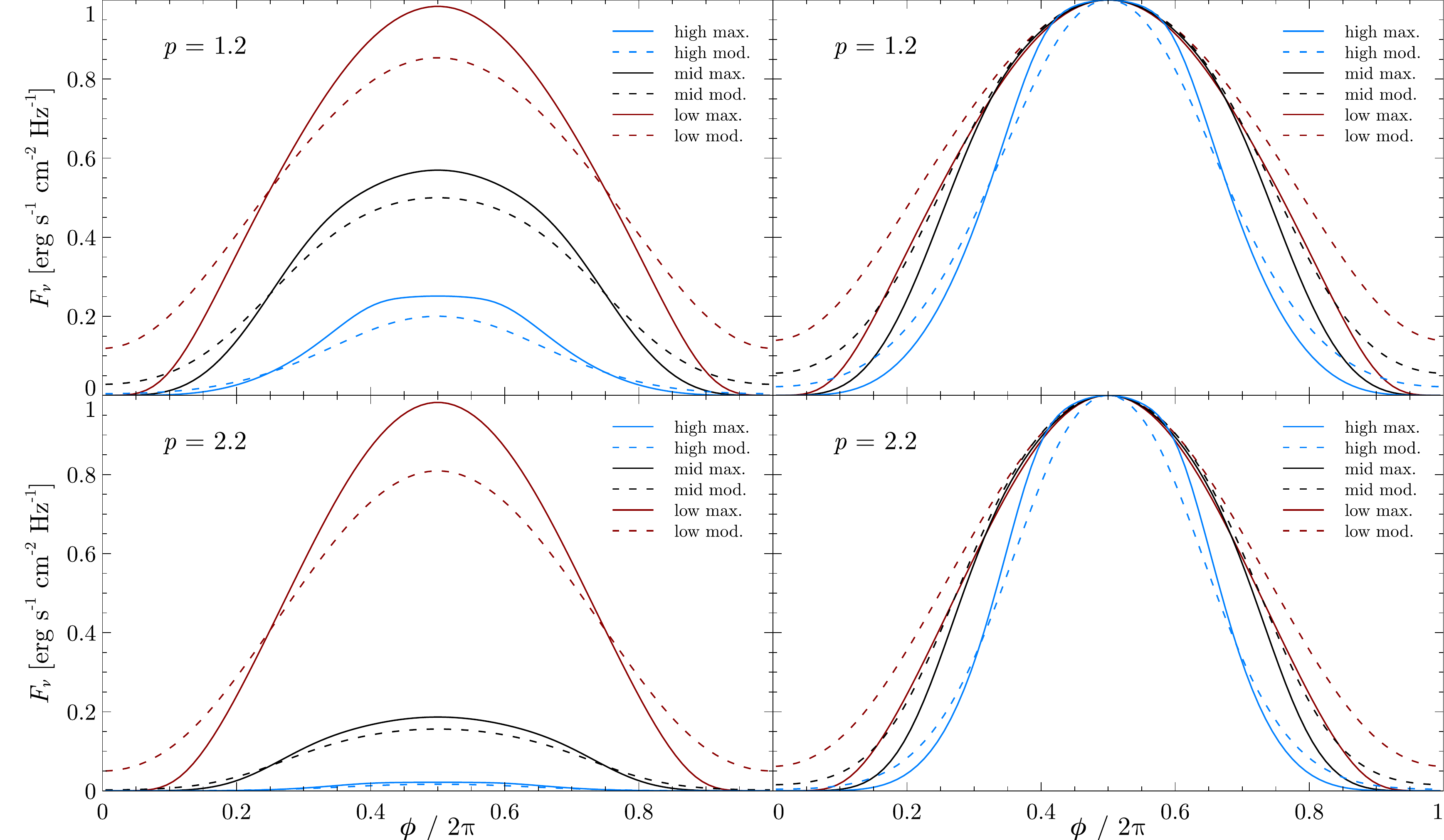}
	\caption{Simulated lightcurves of the Comptonized emission as a function of the neutron star rotational phase $\phi$. We show the lightcurves for two values of the spectral powerlaw index of the pair energy distribution $p\in\{1.2,2.2\}$. Per plot the curves exhibit phase variability in three separate energy bands, i.e. low (260 keV $\leq\varepsilon_1^{\rm low}\leq$ 1.4 MeV), mid (1.4 MeV $<\varepsilon_1^{\rm mid}\leq$ 8.4 MeV), and high (8.4 MeV $<\varepsilon_1^{\rm high}\leq$ 50 MeV), and for two different dependencies of $\theta_0$ on $\phi$. For the latter, $\theta_0$ may vary maximally (moderately) with $\phi$, where $\cos\theta_0 = \cos\phi$ ($\cos_0=1/\sqrt[]{2}\cos\phi$). The left panels show the relative flux of the separate energy bands; from which we see that the pulse amplitude decreases with energy, and decreases stronger for larger $p$.  The right panels show the fluxes normalized by their respective maximum value; from which we see that the anisotropy increases with energy, regardless the value of $p$.}\label{fig:phase}
\end{figure*}

In Fig. \ref{fig:phase} we plot the simulated light curves as a function of the NS rotational phase $\phi$. The plots correspond to two values of $p\in\{1.2,2.2\}$. The solid and dashed curves respectively show a maximum and moderate dependence of the scattering angle on the phase, i.e. consecutively ($\chi=\pi/2,\,\alpha=\pi/2$) and ($\chi=\pi/2,\,\alpha=\pi/4$). Moreover, per plot the averaged flux of three distinct energy bands are shown, in the low (260 keV $\leq\varepsilon_1^{\rm low}\leq$ 1.4 MeV), mid (1.4 MeV $<\varepsilon_1^{\rm mid}\leq$ 8.4 MeV), high (8.4 MeV $<\varepsilon_1^{\rm high}\leq$ 50 MeV) range. 

The left panels show the relative flux of the separate energy bands and the right panels show the fluxes normalized by their respective maximum value. We find that the pulse amplitude decreases with energy and that this effect is stronger at larger values for $p$. Moreover, we find that the anisotropy of the Comptonized emission increases with energy, regardless the value of $p$. The pulsed fraction remains very close to 1.0 in the case of maximum variability, yet does not become much less ($\gtrsim0.85$) in the case of moderate variability. Our analysis shows that the IC mechanism does not isotropize the thermal emission, as suggested by \citet{Boggs2007}. In fact, the emission generated by this IC model remains strongly pulsed in the higher energy bands, contrary to what the data indicate.

\subsection{Discussion and review of assumptions}\label{sec:Discussion and review of assumptions}

\subsubsection{Validity of the Klein-Nishina kernel}\label{sec:Validity of the Klein-Nishina kernel}

In deriving the K-N kernel given by Eq.~(\ref{eq: K-N kernel}) we made use of the approximation that $\gamma\gg1$. Nevertheless, in generating the spectra we set our minimum particle energy at $\gamma_{-}=1$. This however does not only affect the lower energy end of the resultant spectra (below 1 MeV, mainly for smaller values of $\theta_0$), it also weakens the assumption that the Comptonized emission that we observe solely originates from kinematic interactions of seed photons with charged particles that propagate in the direction of the observer. If $\gamma\sim1$ the outgoing photon is not necessarily beamed in the same direction as the motion of the charged particle. Hence, we may observe photons that have interacted with sub relativistic pairs, under a different incident angle than $\theta_0$. This effectively means that the anisotropy will be somewhat reduced for the Comptonized emission at the lower energies, i.e. $\lesssim1$ MeV. Nonetheless, the higher energies are much less affected by this assumption and will remain highly anisotropic as shown in Fig.~\ref{fig:phase}.

\subsubsection{A moving scattering region}

We have assumed that the highly extended corona is stationary. In general however we expect the corona to consist of an outflow; therefore, we can relax the hardness of the electron distribution by assuming that the scattering region is moving mildly relativistically away from the star. In the comoving frame the energy of the thermal photons will be lower, such that the IC scattering is no longer in the K-N regime. Subsequently, we would still require a hard electron distribution with $p \lesssim 1.8$, yet much less hard than the value of $p\lesssim 1.4$ that corresponds to a stationary scattering region with respect to the observer. In the nonstationary case we would expect the hard emission to be as pulsed as the thermal emission, since both populations are boosted with respect to the observer.  One would also expect the optical depth of the scattering region to decrease as it moves away if it also expands.  The similar power-law decay of the flux in the two energy regimes constrains changes in the optical depth.

\subsubsection{IC cooling time \& size of the corona}\label{sec:IC cooling time and size of the corona}

From Eq.~(\ref{eq:energy in scattering region}) we found that the observed high-energy emission can be produced from inverse Compton scattering only if the total scattering region has approximately $3.3 \times 10^{18}~\textrm{ergs cm}^{-2}$ of pairs with $1 < \gamma < 10^2$. Even though there may be higher energy particles, these nonetheless do not contribute much to the emission below 20~MeV. The total energy in the high-energy emission is $\sim10^{42}$~ergs assuming isotropic emission and a distance to the source of $d=8.7$~kpc. This yields a scattering region radius of $r_{\rm sr}\sim4.4\times10^4~R_{\rm NS}$, which exceeds the light cylinder of the source, i.e. $R_{\rm lc}=c\,\Omega_{\rm NS}^{-1}\sim 3.6\times10^4~R_{\rm NS}$. Only if $\gamma_+\gtrsim10^3$, do we find that $r_{\rm sr}< R_{\rm lc}$.

An important question is the cooling time of the scattering population.  If the cooling time is much shorter than the duration of the high-energy excess emission, then the pairs must be continuously replenished. We can estimate the IC cooling time $\tau_{\rm c}^{\rm IC}$ by calculating the total energy column density of the pairs in the scattering region $E_\pm$ over the IC power per unit area $dP_{\rm IC}/dA$, i.e.
\begin{equation}
\tau_{\rm c}^{\rm IC}\sim E_\pm\left(\frac{dP_{\rm IC}}{dA}\right)^{-1}.
\end{equation}
With Eq.~(\ref{eq:isotropic energy distribution of the pairs}) and Eq.~(\ref{eq:energy in scattering region}) we find that we may write
\begin{equation}
E_{\pm}=\frac{m_\pm c^2 K_\pm}{4\pi}\left(\frac{\gamma_+^{-p+2}-\gamma_-^{-p+2}}{2-p}\right).
\end{equation}
The IC power emitted by a single electron is given by 
\begin{equation}\label{eq: IC power}
P_{\rm IC}=\frac{4}{3}\sigma_{\rm T}c\beta^2\gamma^2u_{\rm X},
\end{equation}
where $\sigma_{\rm T}\equiv (8\pi/3)\,r_e^2$ is the Thomson cross section.  We have made a modest approximation by using the Thomson cross section rather than the K-N expression, Eq.~(\ref{eq:K-N cross section}). Accordingly, we may infer the IC power per unit area of the scattering region as follows,
\begin{align}
\frac{dP_{\rm IC}}{dA}&=\int_{\gamma_-}^{\gamma_+}P_{\rm IC}\frac{dN_\pm}{d\gamma}\,d\gamma = \frac{\frac{4}{3}\sigma_{\rm T}c\,u_{\rm X}K_\pm}{4\pi}\int_{\gamma_-}^{\gamma_+}\beta^2\gamma^{-p+2}\,d\gamma,\nonumber\\
&\simeq\frac{\frac{4}{3}\sigma_{\rm T}c\,u_{\rm X}K_\pm}{4\pi}\left(\frac{\gamma_+^{-p+3}-\gamma_-^{-p+3}}{3-p}\right),
\end{align}
where in the last step we assume that the scatterers are relativistic, i.e. $\beta\sim 1$. Consequently, we can write
\begin{equation}
\label{eq:IC cooling equation}
\tau_{\rm c}^{\rm IC}\sim\frac{3m_\pm c}{4\sigma_{\rm T} u_{\rm X}}\left(\frac{3-p}{2-p}\right)\frac{1}{\gamma_-}\left[\frac{\left(\frac{\gamma_+}{\gamma_-}\right)^{-p+2}-1}{\left(\frac{\gamma_+}{\gamma_-}\right)^{-p+3}-1}\right].
\end{equation}
The energy density of the incident X-ray emission can be estimated as
\begin{equation}\label{eq: energy density soft emission}
u_{\rm X}=\frac{L_{\rm X}}{\pi c R_{\rm NS}^2}\left(\frac{r_{\rm sr}}{R_{\rm NS}}\right)^{-2},
\end{equation}
 where $L_{\rm X}$ denotes the (isotropic) X-ray flux at $R_{\rm NS}$, and $r_{\rm sr}$ represents the radial height of the scattering region.

Assuming $\gamma_+ \gg \gamma_-$, and adopting typical values for $p=1.2$, $L_{\rm X}\sim10^{42}$ erg s$^{-1}$, and $r_{\rm SR}\sim10^4\,R_{\rm NS}$, we obtain an IC cooling time of, 
\begin{equation}\label{eq: IC cooling 10^4}
\tau_{\rm c}^{\rm IC}\sim\frac{7\times 10^{-4}}{\gamma_+} \left ( \frac{r_{\rm sr}}{10^4 R_{\rm NS}} \right)^2 \text{ s },
\end{equation}
{\em i.e.} the pairs cool very rapidly through IC scattering.

Of course this cooling mechanism would only affect the pairs that are illuminated by the beam of soft photons. We expect the pairs in these illuminated regions to be replenished high-energy particles in neighboring regions. Observations of eruptive solar flares and magnetic reconnection models
indicate that the total energy in this highly extended corona  could be
comparable to the energy released during the initial spike of the GF
{\citep{2014ApJ...783L..21S,2015SoPh..290.3425J,2016MNRAS.462...48S}.
} Therefore, there should be sufficient total energy in the corona to power the late high-energy emission as long as other cooling mechanisms do not operate. 

\subsubsection{Synchrotron cooling time \& lower limit to the size of the corona}\label{sec:Synchrotron cooling time and lower limit to the size of the corona}

Synchrotron emission might however cool the corona over the timescale of the high-energy emission.  If we insist that the synchrotron cooling time is longer than 100~s, we can obtain a lower limit on the size of the corona and the energy contained within it. We can estimate the synchrotron cooling time $\tau_c^{\rm syn}$ by taking Eq.~(\ref{eq: IC power}) and substituting $u_X\to u_B$, where the latter is the magnetic energy density given by,
\begin{equation}
u_B=\frac{B^2(r)}{8\pi}=\frac{B_0^2}{8\pi}\left(\frac{r}{R_{\rm NS}}\right)^{-6},
\end{equation}
assuming that the post-GF magnetic field resembles a dipole. We end up with the following estimate,
\begin{equation}\label{eq: synchrotron cooling time 10^4}
\tau_{\rm c}^{\rm syn} \sim \frac{4 \times 10^2}{\gamma_+} \left ( \frac{r_{\rm sr}}{10^4 R_{\rm NS}} \right )^{6} {\rm s},
\end{equation}
where we adopted a surface magnetic field strength of $B_0\simeq2\times10^{15}$ G \citep{Nakagawa2009}. Incidentally, the synchrotron cooling time will decrease in the case of a globally twisted magnetic field \citep{Thompson2002}.

The synchrotron cooling time of pairs with $\gamma_+\geq10^2$ can be stretched beyond 100~s only if the scattering region is larger than $1.7\times 10^{4}~R_{\rm NS}$. This means that the initial cloud of relativistic particles, that was presumably created near to the star during the initial spike of the GF, must expand by a factor more than $\sim10^4$ without cooling due to inverse Compton scattering, synchrotron radiation or adiabatic expansion. The adiabatic effects alone reduce the Lorentz factor of the pairs by a factor of $10^4$, dramatically increasing the energy requirements. Accounting for adiabatic losses alone would require an initial energy in relativistic particles at the reconnection event of $\sim 10^{46}$~ergs, as much as the energy released in the GF itself.  We must therefore conclude that the high-energy emission results from pairs that are continuously accelerated by the outflow rather than from pairs excited by the initial GF.

\section{Local Flare Model: Shock Acceleration}

\begin{figure}
\centering
\includegraphics[width=\columnwidth,clip,trim=2in 5in 2in 2in]{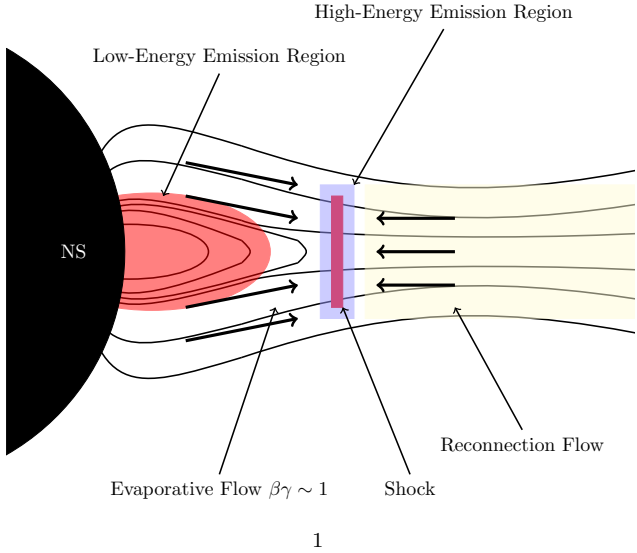}
1\caption{Local flare model in which the outflow generated near the surface forms a shock that accelerates particles and perhaps generates a magnetic field as well. The thickness of the high-energy emission region is limited by the cooling time.}
\label{fig:flare_model}

\end{figure}

The initial spike of the GF is expected to generate copious pairs and a large initial outflow.  Outflows of material must persist throughout the tail of the giant flare however, in order to collimate the low energy emission and generate the prominent rotational pulsations \citep{vanPutten2016}.  These outflows are driven by the super-Eddington emission from the base of the trapped fireballs. We find that both the soft and hard photons decrease at a similar rate, i.e. the soft emission flux also decays as $f(t)\propto t^{-0.76}$ (see Section~\ref{sec:temporal and spectral properties}). This indicates a possible connection between the particle injection rate and the soft emission that originates from the outflow. We present a local emission model, inspired by the CSHKP model \citep[after][]{Carmichael1964,Sturrock1968,Hirayama1974,Kopp1976} for solar flares \citep[e.g.][for a review]{Shibata2011}, in which the particles in the outflowing material can be accelerated sufficiently to replenish continuously the population of relativistic particles. This can be achieved if the outflow from near the surface of the NS collides with the exhaust of the reconnection region as depicted in Fig.\ref{fig:flare_model}, forming a shock that accelerates particles. Significant advances in the study of relativistic shocks have recently been made both theoretically and numerically through Particle-In-Cell (PIC) simulations. In particular their ability for particle acceleration and generating magnetic fields in the upstream region has been investigated \citep{Sironi2015}. It has been determined that the physics of relativistic shocks is strongly dependent on the value of the magnetization parameter $\sigma_{\rm m}\equiv2u_B/u_\pm$. Fundamentally, the conditions for the Fermi acceleration mechanism to operate are constrained by the orientation of the field with respect to the upstream flow in strongly magnetized shocks ($\sigma_{\rm m}\gtrsim10^{-3}$ for pair dominated flows). Only for shocks where the orientation of the upstream magnetic field is near-parallel to the shock normal, can efficient particle acceleration occur. In weakly or unmagnetized shocks ($\sigma_{\rm m}\lesssim10^{-3}$ for pair dominated flows) however, accelerated particles are more easily able to flow into the upstream region and excite electromagnetic plasma instabilities, which in turn produce magnetic structures that facilitate the Fermi acceleration mechanism. 

\subsection{Inverse-Compton Scattering}
\label{sec:ICscattering}
We first analyze the case in which the high-energy emission is produced by IC scattering of the low-energy photons by the particles accelerated in the shock. In this case, we can compare the fluence of the high-energy emission to that of the low-energy emission to obtain an estimate of the total energy in relativistic particles up to a Lorentz factor of $10^2$. Although higher energy particles are likely to be present, they do not contribute to the emission $< 20$ MeV. The total energy column density in the scatterers was determined in the previous Section [Eq.~(\ref{eq:energy in scattering region})] to be
\begin{equation}\label{eq: relation energy column density and energy density}
E_\pm = u_\pm l = 3.3 \times 10^{18}\,\text{ergs~cm}^{-2},
\end{equation}
where here $l$ is the thickness of the emission layer. 

Meanwhile, the luminosity of the high-energy emission is about $3.8 \times 10^{40}\, \textrm{ergs~s}^{-1}$.  Beyond about 350 stellar radii, inverse Compton cooling will dominate over synchrotron cooling, assuming that all of the magnetic field originates from the star. Rescaling Eq.~(\ref{eq: IC cooling 10^4}) we find that in this case
\begin{equation}
\tau_{\rm c}^{\rm IC}\sim\frac{8\times 10^{-7}}{\gamma_+} \left ( \frac{r_{\rm sr}}{350\,R_{\rm NS}} \right)^2 \text{ s },
\end{equation}
and with $\gamma_+=10^2$, we estimate a flux from the pairs at the source of 
\begin{equation}
F_\pm\sim4.1 \times 10^{26}   \left ( \frac{r_{\rm sr}}{350\,R_{\rm NS}} \right)^{-2}~\text{cm}^{-2}\,\text{erg~s}^{-1}.
\end{equation}
Combining this with the aforementioned luminosity yields the surface area of the scatterers
\begin{equation}
\Sigma_\textrm{tot} \sim 9 \times 10^{13}  \left ( \frac{r_{\rm sr}}{350 R_{\rm\,NS}} \right)^{2}~\text{cm}^2.
\end{equation}
The combined cross section of scatterers can only subtend a tiny fraction of the magnetosphere, i.e. roughly $6\times 10^{-5}$ of the sphere.\footnote{An equivalent way to look at this is to estimate the scattering optical depth. The typical photon increases its energy upon scattering by a factor of about 100, and the total luminosity in the soft component is 100 times larger than in the hard component, yielding a scattering optical depth of about $10^{-4}$.}  

We can also obtain an estimate of the minimal size of the scattering region by determining how far a relativistic particle can travel before cooling. The typical distance between scattering events is
\begin{equation}
\lambda_{\rm mfp} = \frac{\langle\epsilon_{\rm bb}\rangle}{\sigma_{\rm T} u_{\rm X}},
\end{equation}
where $\langle\epsilon_{\rm bb}\rangle\sim 27$ keV is the mean energy of a soft photon, for $k_{\rm B}T_{\rm bb}=10$ keV. The typical number of scattering events is 
\begin{equation}
N_{\rm scatt}=\frac{c \tau_c^{\rm IC}}{\lambda_{\rm mfp}} 
\sim 1.7\,\frac{m_\pm c^2}{\gamma_+ \langle\epsilon_{\rm bb}\rangle},
\end{equation}
with the relative change in the particle momentum of
\begin{equation}
\frac{\Delta p}{p}  
\sim \frac{\gamma_+ \langle\epsilon_{\rm bb}\rangle}{m_\pm c^2},
\end{equation}
per scattering event. The particle essentially travels a straight path as it loses energy, so the minimum size of the scattering region is determined by the cooling time 
\begin{equation}
c \tau_c^{\rm IC} \sim \frac{2.4\times10^4}{\gamma_+} \left ( \frac{r_{\rm sr}}{350 R_{\rm NS}} \right )^2~\text{cm}  = \frac{6.8\times 10^{-5} r_{\rm sr}}{\gamma_+} \left (\frac{r_{\rm sr}}{350 R_{\rm NS}}\right).
\label{eq:iccoollength}
\end{equation}

Furthermore the spectrum of scattered photons, even if the maximum Lorentz factor $\gamma_+$ is large, will only extend to an energy
\begin{equation}
\epsilon_\textrm{max} = \gamma_{\rm max} m_\pm c^2 = \gamma^2_{\rm max} \langle\epsilon_{\rm bb}\rangle = \frac{m_\pm^2 c^4}{\langle\epsilon_{\rm bb}\rangle} \sim 10~\textrm{MeV}
\label{eq:emaxic}
\end{equation}
because above this energy the K-N cross section becomes important.

In general when a shock accelerates particles, it also generates a magnetic field such that $u_\pm \simeq u_B$, where here the latter is the energy density of magnetic field generated in the shock \citep{Sironi2015,2016RPPh...79d6901M}. Because we are examining a model where IC scattering dominates the high-energy radiation, we must also insist that $u_{\rm X} > u_B$ and accordingly $u_{\rm X}>u_\pm$. Subsequently, we find that the thickness of the high-energy emission layer $l$ must satisfy
\begin{equation}
l > \frac{u_\pm l}{u_{\rm X}}= 4\times10^4 \left (\frac{r_{\rm sr}}{350\,R_{\rm NS}} \right )^2~\text{cm},
\end{equation}
where we made use of Eq.~(\ref{eq: energy density soft emission}) and Eq.~(\ref{eq: relation energy column density and energy density}). For $\gamma_+ \sim 10^2$ which is necessary to account for the high-energy emission, this exceeds the cooling length by a factor of more than 100, creating difficulties for the IC model, if the particles are accelerated in a region thinner than $l$. If the region containing the relativistic particles is much thinner than this and magnetic field is also generated in the shock at a similar rate to the relativistic particles, synchrotron emission will dominate over IC.

Furthermore, given the analysis in Section~\ref{sec:icmodel}, we expect IC scattering to yield a cutoff in the high-energy emission around $20-50$~MeV and to result in pulsed high-energy emission. The analysis of the \emph{RHESSI} data yields marginal evidence that the emission is at most weakly pulsed and there is no cutoff to the high-energy emission below 50~MeV.

\subsection{Synchrotron radiation} 

We will now examine an alternative picture in which the high-energy emission results from synchrotron radiation.  The magnetic field can be due to the star or generated in the shock itself \citep{Sironi2015,2016RPPh...79d6901M}. In both cases we will find that the cutoff in the high-energy emission may be small if the emission region is close to the star, but it is typically much larger if the shock is further away. Furthermore, if the magnetic field is generated in the shock itself, we expect it to be chaotic and the high-energy emission may be very weakly pulsed.

If the emission region is either close to the star (within 350~stellar radii) or thinner than a few hundred meters, synchrotron emission will indeed dominate over IC emission under the further assumption that the field strength decreases as a dipole from a surface field of about $2 \times 10^{15}$~G. The pulsed fraction of the high-energy emission in this case will be low if the local magnetic field is chaotic. In principle, the magnetic field could become highly distorted either through reconnection or shocks. The field from the star at a distance of 350 $R_{\rm NS}$ is strong, but at about $5 \times 10^7$~G it is less than the quantum critical field ($B_{\rm qed}\simeq4.4\times10^{13}$ G), so synchrotron emission should proceed classically and extend into the MeV regime for electrons or positrons with $\gamma \sim 10^3$. 

Rescaling Eq.~(\ref{eq: synchrotron cooling time 10^4}), we note that at 350 $R_{\rm NS}$ the synchrotron cooling time is also short, i.e. $8\times10^{-7}\gamma_+^{-1}$s, which also requires that particles are continually accelerated to high energies; we therefore expect the high-energy flux to reflect the instantaneous particle injection rate. In the synchrotron picture, assuming that the particles are accelerated in a shock where the mildly relativistic outflow is halted, we would expect the particle injection rate to decrease in time as the soft flux does, and therefore expect the high-energy emission to decrease at the same rate as the soft emission.

Another observational feature that can pinpoint whether the emission is IC or synchrotron radiation is the cutoff energy. The \emph{RHESSI} data do not indicate a cutoff energy $< 50$ MeV. For synchrotron emission the cutoff depends on the strength of the magnetic field in the emission region; specifically, the maximum Lorentz factor $\gamma_{\rm co}$ of the pairs that emit classical synchrotron radiation is given by
\begin{equation}
\epsilon_{\rm co} = \gamma_{\rm co} m_\pm c^2 = \gamma_{\rm co}^2 \epsilon_{\rm cyc} = \frac{m_\pm^2 c^4}{\epsilon_{\rm cyc}} 
\end{equation}
where $\epsilon_{\rm cyc}=\hbar eB/(m_\pm c)$ is the non-relativistic cyclotron energy. For pairs with larger energy, the emission must be treated quantum mechanically. Therefore, there is a cutoff in the photon spectrum at this energy.

In order to estimate the cutoff for synchrotron emission, we have to estimate the magnetic field strength in the shock region. For this reason, we have to make some assumptions. The first one is obvious: we are analyzing the case in which synchrotron dominates over IC, so we want $u_B \gtrsim u_{\rm X}$, where $u_B$ is the energy density of the magnetic field generated by the shock. We will consider $u_B = a \, u_{\rm X}$, where $a \gtrsim 1$ is a boost parameter that depends on the distance of the shock region from the star.

The second constraint on the energy in the magnetic field is given by the assumption that the energy in the pairs accelerated by the shock and the energy in the magnetic field generated by the shock have to be the same order of magnitude, i.e. $u_\pm\sim u_B$. We can estimate the energy in the pairs by multiplying the luminosity of the high-energy emission ($L_{\rm Hi}\sim3.8\times10^{40}$ ergs s$^{-1}$) by the synchrotron cooling time $\tau_{\rm c}^{\rm syn}$, which in turn is given by Eq.~(\ref{eq:IC cooling equation}) and substituting $u_X\to u_B$ and setting $p=1.8$. 
In Eq.~(\ref{eq:IC cooling equation}), $\gamma_+$ and $\gamma_-$ are defined as the maximum and minimum Lorentz factors for electrons emitting at the upper and lower limit of \emph{RHESSI}'s band respectively:
\begin{equation}\label{eq: e_phot}
\epsilon_+ \approx 0.29 \frac{3}{2} \gamma_+^2 \epsilon_{\rm cyc} \sin \alpha
\end{equation}
where $\epsilon_+ = 15$ MeV is the upper limit of \emph{RHESSI}'s energy band, where we take $\sin \alpha\approx 0.5$ and the factor $0.29$ is where the first synchrotron function $F(x)$ peaks \citep{Rybicki1986}. $\gamma_-$ is defined in the same way, with $\epsilon_- = 0.4$ MeV  (the lower limit of \emph{RHESSI}'s high-energy energy band). The value of the limiting Lorentz factors depends on the strength of the magnetic field.
The other missing piece of information is the volume of the emission region. The minimum thickness of the region is given by how far an electron moves in one cooling time. Assuming random walk, this is given by
\begin{equation}\label{eq: minimum thickness emission region relation}
l \gtrsim \sqrt{c\tau_c^{\rm syn}\rho}
\end{equation}
where $\rho = \gamma_- m_\pm c^2/(eB)$ is the Larmor radius.

We then indicate the surface area of the region by $q r_{\rm s}^2$, where $r_{\rm s}$ is the distance of the shock region from the star and $q < 1$.
Our assumption that the energy in the magnetic field equals that in the pairs can then be translated in the following expression
\begin{equation}\label{eq: energy balance}
\frac{L_{\rm Hi} \tau_c^{\rm syn}}{u_{\rm B} l q r_{\rm s}^2} \simeq 1.
\end{equation}
Accordingly we find,
\begin{equation}
u_B \simeq 9 \times 10^{13}\left( \frac{q}{0.8} \right)^{-1} \left( \frac{r_{\rm s}}{350\,R_{\rm NS}} \right)^{-2}~\text{ergs cm}^{-3},
\end{equation}
which corresponds to a magnetic field of
\begin{equation}
B \simeq 4.8 \times 10^{7} \left(\frac{q}{0.8} \right)^{-1/2} \left( \frac{r_{\rm s}}{350\,R_{\rm NS}} \right)^{-1}~\text{G}.
\end{equation}
This is given by imposing a minimum boost on the magnetic energy density compared to the energy density in the photons: $u_B = a \, u_{\rm X}$, that does not depend on $r_{\rm s}$
\begin{equation}
a \sim 1 \, \left(\frac{q}{0.8} \right)^{-1}.
\end{equation}
Since in order for the emission to be dominated by synchrotron, $a$ has to be greater than 1. Accordingly, we expect the area of the emission region to be smaller ($q < 0.8$). With these values, the minimum thickness of the emission region [Eq.~(\ref{eq: minimum thickness emission region relation})] becomes
\begin{equation}
l \simeq 0.35 \left(\frac{q}{0.8} \right)^{3/4} \left( \frac{r_{\rm s}}{350\,R_{\rm NS}} \right)^{3/2} {\rm cm}.
\end{equation}
This is assuming a zero thickness for the shock itself, so the thickness of the emission region is determined by the cooling length. Of course, there is still some arbitrariness in these estimates since the shock region can be thicker; equivalent to increasing $a$ and decreasing the energy in the field, or have a smaller area (lower $q$), increasing the energy in the magnetic field. However, this picture is robust against changes in the parameters. For example, one could reduce the distance of the shock region from the star to $35 R_{\rm NS}$ and still have the field produced in the shock itself exceed the stellar dipole field, for $q<8 \times 10^{-5}$, corresponding to a value of the boost parameter $a=10^4$. If the shock region was much larger than this, then one would expect that the dipolar magnetic field may dominate over the one generated in the shock. In this case, synchrotron emission could still account for the high-energy photons, but it would be harder to explain a low pulse fraction with the more organized field configuration.

We find that the maximum Lorentz factor and, consequently, the cutoff energy, increase with the distance to the star:
\begin{equation}
\gamma_{\rm co} \simeq 9.3 \times 10^5  \left ( \frac{q}{0.8} \right )^{1/2} \left( \frac{r_{\rm s}}{350\,R_{\rm NS}} \right)  
\end{equation}
\begin{equation}
\epsilon_{\rm co} \simeq 473 \left ( \frac{q}{0.8} \right )^{1/2} \left( \frac{r_{\rm s}}{350\,R_{\rm NS}} \right)  \,   {\rm GeV}
\end{equation}
For example, in the case where the shock region is 10 times closer and much smaller so the shock-generated field still dominates over that of the star, $\epsilon_{\rm co}$ is $10^3$ times smaller at about 0.5~GeV. Measuring a cutoff in the Fermi energy band would point in the direction of a synchrotron emission process as the source of the high-energy emission and would give us information on the geometry of the emission region.

\section{Fluence prediction for Fermi Gamma-Ray Space Telescope}\label{sec:fluence prediction for current missions}

\begin{table*}
\caption{Predicted number of counts to be observed with the Fermi detectors (\emph{LAT} and \emph{GBM}) in the event of a giant flare comparable to the 2004 December 27 burst from SGR 1806--20. Estimates were made for two distinct inferred photon spectral models: a power law (PL: with $\Gamma=1.38$) and inverse Compton model (IC model: with $p=1.2$, $\theta_0=0.59\pi$, and $\gamma=10^6$). For comparison, \emph{RHESSI} observed the latter at $5^\circ$ off-axis and recorded $\sim2.1\times10^3$ counts in the energy range $0.4-15$ MeV integrated over $\Delta t_{\rm spec}=20.35$ s. All reported numbers of counts are similarly the result of integrations over $\Delta t_{\rm spec}$.}
\centering
\begin{tabular}{l|ccccc}
\hline
GRB & $\zeta_{LAT}$ & Model & $N_{LAT}/10^3$ & $N_{GBM}^{\rm b}/10^3$\\
identifier$^{\rm a}$&&& $(80-10^4)$ MeV & $(0.4-38)$ MeV$^{\rm c}$ \\
\hline
\hline
090510A & $14^\circ$ &PL& $22.0\pm0.7$ & $4.8\pm0.9$\\
& &IC& $2.0\pm0.2$ & $4.5\pm0.8$\\
160509A & $32^\circ$ &PL& $18.8\pm0.8$ & $4.9\pm0.9$\\
& &IC& $1.8\pm0.2$ & $4.5\pm0.8$\\
110721A & $41^\circ$ &PL& $16.2\pm0.6$ & $4.5\pm0.8$\\
& &IC& $1.5\pm0.2$ & $4.4\pm0.8$\\
080916C & $49^\circ$ &PL& $14.4\pm0.7$ & $4.7\pm0.9$\\
& &IC& $1.3\pm0.2$ & $4.7\pm0.9$\\
150210A & $54^\circ$ &PL& $10.4\pm0.5$ & $4.1\pm0.8$\\
& &IC& $0.8\pm0.1$ & $4.1\pm0.8$\\
\hline
\end{tabular}\begin{flushleft}{\small $^{\rm a}$These denote the names of the $\gamma$-ray bursts (GRBs) from which we used the instrument response files to generate the fake count spectra; obtained from the Fermi \emph{GBM}/\emph{LAT} GRB catalog: \url{http://fermi.gsfc.nasa.gov/ssc/observations/types/grbs/lat_grbs/index.php}. $^{\rm b}$These values represent the sum total of the simulated counts in the 3 NaI and BGO detectors where the GRB signal was the brightest. $^{\rm c}$The lower energy limit of the \emph{GBM} has been set to the lower limit of the observed nonthermal emission.}\end{flushleft}
\label{tab:predict results}
\end{table*}

\begin{figure}
	\centering
		\includegraphics[width=0.45 \textwidth]{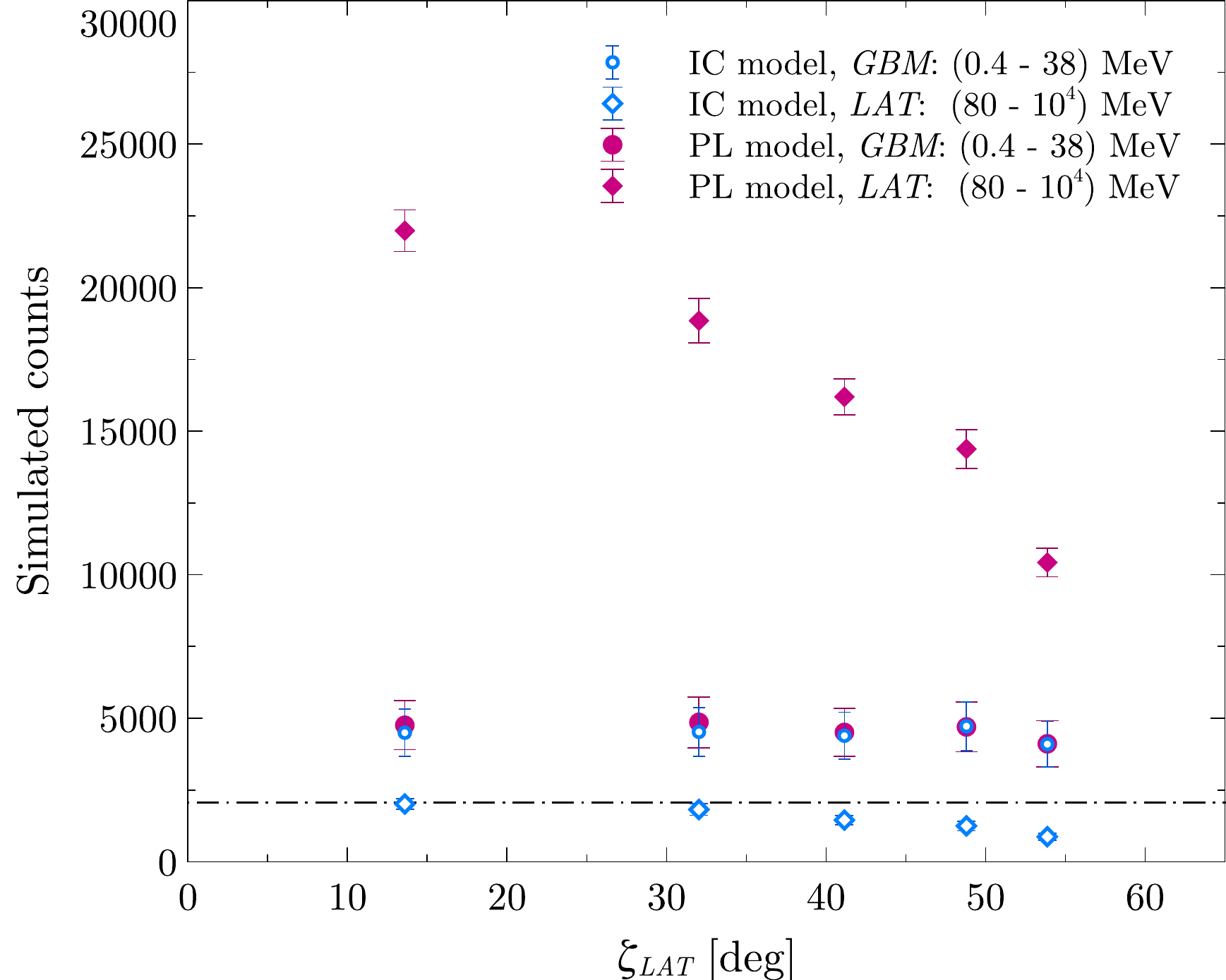}
	\caption{Simulated number of counts versus angle of Fermi \emph{LAT} boresight $\zeta_{LAT}$. We compare the results for the predicted number of counts in the separate Fermi instruments, i.e. the \emph{GBM} and the \emph{LAT}, based on the powerlaw (PL) and inverse Compton (IC) model. By far the most counts are expected to be detected with the \emph{LAT} if the PL extends up to $10^4$ MeV. Moreover, we find that the predicted number of counts of the \emph{LAT} are dependent on $\zeta_{LAT}$ and decrease with increasing incident angle, whereas those for the \emph{GBM} remain roughly constant. For comparison the number of counts detected by \emph{RHESSI} is shown by the horizontal dash-dotted line at $\sim2.1\times10^3$.}\label{fig:counts_v_angle}
\end{figure}

Fermi is designed to observe high-energy astrophysical phenomena over a large energy bandwidth with its two main instruments, the \emph{GBM}, with energy bandpass from practically 10 keV to 38 MeV, and the \emph{LAT}, covering an energy range $\sim80-10^4$ MeV. There has been no magnetar GF during the lifetime of Fermi, which has been operational since 2008. It is currently the best instrument available to study the highest energy components of the emission should a magnetar GF occur and previous recurrence times suggest one could be imminent. 

Here we aim to predict the total number of counts that Fermi would detect in the event of a magnetar GF, similar to the 2004 December 27 event. With knowledge of the incident photon spectrum and flux as inferred from the \emph{RHESSI} data, we can fold the obtained photon models (shown in Fig.~\ref{fig:IC_spec_fit}) through the appropriate response (see Appendix \ref{sec:instrument responses}) using \texttt{fakeit} in \texttt{XSPEC} to find the count energy distribution that would be recorded by the respective instruments. The inferred photon spectral models, i.e. a PL (with $\Gamma=1.38$, appropriate for synchrotron emission) and an IC model (with $p=1.2$,  $\theta_0=0.59\pi$, and $\gamma_+=10^6$), are based on a fit over the energy range $0.4-15$ MeV. We know that below 0.4 MeV the spectrum deviates from the deduced PL and IC spectrum. However, in the case of the PL spectrum we cannot determine the cut off energy from the \emph{RHESSI} data; for the predictions we will assume that the PL extends into the \emph{LAT} bandwidth. \citet{Boggs2007} note that they retrieve a slightly worse fit to the \emph{RHESSI} count spectral data, when they introduce an exponential cut off at $> 50$ MeV, compared to the simple PL model. The simulated count data will provide us with an estimate of the number of counts that may be observed in the event of a GF. We will adopt the same amount of integration/observation time that was used in generating the \emph{RHESSI} count spectrum, i.e. $\Delta t_{\rm spec}=20.35$ s.

We obtained response matrices for the respective instruments from the online Fermi \emph{GBM} and \emph{LAT} $\gamma$-ray burst (GRB) catalogs\footnote{Online Fermi \emph{GBM} and \emph{LAT} $\gamma$-ray burst catalog: \url{http://fermi.gsfc.nasa.gov/ssc/observations/types/grbs/lat_grbs/index.php}}. We selected the instrument response files that were used in the spectral analysis of 5 bright GRBs, which respectively occurred at an angle $\zeta_{LAT}$ to the \emph{LAT} boresight - see Table~\ref{tab:predict results}. 

The \emph{GBM} detector is composed of 12 Thallium Sodium Iodide crystals (NaI) and 2 Bismuth Germanate crystals (BGO) \citep[e.g.][]{Meegan2009,Bissaldi2009} that provides full coverage of the non occulted sky.  We selected the response functions from 3 of the NaI and the BGO detectors where the signal was most prominent. We assume that in the event of a GF we can take the total sum of counts recorded in these 4 detectors as an estimate for the counts that the \emph{GBM} would detect. 

Depending on data from the \emph{GBM} or the \emph{LAT}, the \emph{GBM} may send a request for autonomous repointing (ARR) of the spacecraft. The \emph{LAT} changes the observing mode to monitor the location of the transient in or near its FoV. After a pre-determined time (generally $\sim2.5$ hours) the spacecraft will return to the scheduled mode. An ARR is performed within approximately $\sim10$ s \citep{Meegan2009}. Slewing of the Fermi spacecraft changes the photon angle of incidence $\zeta_{LAT}$, which alters the instruments response function. We can estimate the advantageous effect of slewing towards the source by looking at the responses of GRBs that were observed at different incident angles, where a lower $\zeta_{LAT}$ results in a higher number of observed counts with the \emph{LAT} - see Figure~\ref{fig:counts_v_angle}.

The results are summarized in Table~\ref{tab:predict results}. Compared to the number of high-energy counts observed with \emph{RHESSI}, i.e.  $\sim2.1\times10^3$ counts, we find that the \emph{LAT} would roughly observe an order of magnitude more counts in the case of a PL extending up to $10^4$ MeV and a similar or less amount in the case of a IC photon spectrum. The total sum of the 3 NaI and one BGO detector of the \emph{GBM} instrument would detect roughly $2-2.5$ times the number of counts that \emph{RHESSI} observed for both models. Setting the number of counts against the incident angle $\zeta_{LAT}$, as shown in Fig.~\ref{fig:counts_v_angle}, we find that as expected for the \emph{LAT} the number of counts decreases with increasing opening angle, whereas those for the \emph{GBM} remain approximately constant.

In any case, we expect that the Fermi data will by virtue of the increase in the observed number of counts, allow more stringent tests of the physical mechanism that generates the high energy emission, in particular the IC model.

\section{Conclusion}

In this work we aimed to reproduce the results of the \emph{RHESSI} data analysis done by \citep{Boggs2007} on the high energy excess emission observed during the giant flare of SGR 1806--20 on 2004 December 27. Moreover, we set out to study the proposed underlying physical mechanism that may produce such high energy emission, i.e. IC scattering in a highly extended relativistic pair corona, and investigate the constraints it places on such extraordinary phenomena. 

We were able to confirm the earlier results of \citep{Boggs2007} regarding the general properties of the high-energy emission. Crucial however to testing the aforementioned model was to determine whether the high energy component ($> 250$ keV) could be the result of a pulsed source as is apparent for the low energy emission ($< 250$ keV). Using more detailed statistical analysis/robust method we have shown that the existing data from \emph{RHESSI} are consistent with a model where the high-energy emission is not pulsed, but the data are only weakly constraining.

Subsequently, we have shown that IC scattering of a highly anisotropic beam of soft radiation emerging from a trapped fireball could generate this high energy emission given a suitable population of scatterers.  The IC model fits the observed counts best for $p = 1-1.4$, i.e. for a scattering population that has been accelerated (very) efficiently. The origin of the scattering population however remains an open question.   If the scattering population is moving mildly relativistically or if the emission is produce by synchrotron radiation, the particle distribution is a bit softer $p\approx 1.8$.  In either case, the relativistic particles have to be replenished continuously to account for the emission.   The current data are insufficient to decide the issue; better constraints on the time decay of the hard emission and a determination of whether it is pulsed or whether it cuts off below 50~MeV would help to decide the issue.  We find that according to the IC model the outgoing high energy emission will not be isotropized in the process, in fact a significant pulsed fraction is to be expected for the Comptonized emission if this is also observed for the low energy emission.  Furthermore in the IC model we also expect the emission to cut off below 50~MeV.  

Next, we examined a local flare model where synchrotron emission dominates the hard band.  Synchrotron will dominate if the emission region is within about 350 stellar radii or if it is thinner than a few hundred metres.  If the magnetic field is generated in a shock along with the relativistic pairs, the synchrotron emission may only be weakly pusled and can continue into the GeV regime without a cutoff. 

Fermi \emph{LAT/GBM} is very well placed to detect the next giant flare. We have estimated that, depending on the model for the production of the high energy emission, Fermi \emph{GBM} and \emph{LAT} may respectively detect approximately $2-2.5$ times and $\sim1-10$ times the number of counts that \emph{RHESSI} observed in the event of a similar giant flare to the one from SGR 1806--20. With better observations, i.e. more observed counts and in a broader high-energy range, we may be able to observe and analyse in more detail properties of the Comptonized or synchrotron emission, thereby placing tighter constraints on the models for the high-energy emission of magnetar giant flares.

\section*{Acknowledgements}
CE and ALW acknowledge support from NOVA, the Dutch Top Research School for Astronomy. EB acknowledges the Italian `Fondo di Sviluppo e Coesione 2007-2013 - APQ Ricerca Regione Puglia - Future In Research'. CO, IC, and JSH acknowledge support from the Natural Sciences and Engineering Research Council of Canada, the Canadian Foundation for Innovation and the British Columbia Knowledge Development Fund. CO also acknowledges support from the Japanese Ministry of Education, Culture, Sports, Science and Technology. We would like to thank Eric Bellm, Steven Boggs, Gordon Hurford, and David Smith for helpful discussions and advice relating to the \emph{RHESSI} giant flare data. 

\bibliographystyle{mnras}
\bibliography{biblio}

\clearpage 
\appendix

\section{Giant Flare Pulse Model}
\label{sec:pulsemodel}

\begin{table*}
\renewcommand{\arraystretch}{1.3}
\footnotesize
\caption{Overview of the priors used in the Bayesian model}
\begin{threeparttable} 
\begin{tabularx}{18cm}{p{6cm}p{6cm}p{6cm}}
\toprule
\bf{Parameter} & \bf{Meaning} & \bf{Probability Distribution} \\ \midrule
{\it Shared Parameters}  \\ \midrule 
$\Gamma_i$	& power law index &   $\mathrm{Uniform}(0,4)$ \\
$\log{A_\mathrm{PL}}$			& power law amplitude & $\mathrm{Uniform}(-10,10)$ \\
$C$ & constant parameter scaling & $\mathrm{Normal}(1, 0.5)$ \\
$w_i$ & Poisson noise amplitude &  $\mathrm{truncated_Normal}(0, 0.1, 0.0, \infty)^{\mathrm{\emph{a}}}$ \\ \midrule
{\it Variable pulse fraction model $M_2$} \\ \midrule
$a_i$ & scaling offsets & $\mathrm{Uniform}(-2, 2)$\\ 
\bottomrule
\end{tabularx}
   \begin{tablenotes}
      \item{An overview over the model parameters with their respective prior probability distributions.}
      \item[\emph{a}]{Truncated normal distribution with lower bound and no upper bound.}
\end{tablenotes}

\end{threeparttable}
\label{tab:priors}
\end{table*}

We use Bayes rule to build a posterior probability distribution for the PSDs of all four energy bins simultaneously:

\begin{equation}
p(\pars \given \dd, \hypers, M) = \frac{p(\dd \given \pars, \hypers, M) p(\pars \given \hypers, M)}{p(\dd \given \hypers, M)} \; ,
\end{equation}

\noindent where $\pars$ denotes a vector of all model parameters, $\dd = \{D_i\}_{i=1}^N$ is the set of power spectra extracted from each energy bin, $\hypers$ denotes a set of hyper-parameters for the prior distributions on $\theta$ (see Section \ref{sec:priors} below for details), and $M$ denotes any additional model assumption. 

In addition, we denote $\likelihood(\theta) = p(\dd \given \pars, \hypers M)$ as the likelihood, i.e.\ the probability of the data given the model (parameters), $\pi(\pars) = p(\pars \given \hypers, M)$ as the prior probability of the parameters $\pars$,  and  $p(\dd \given \hypers, M) = \int_{\Omega} p(\dd \given \pars, \hypers, M) p(\pars \given \hypers, M) d\Omega $ is the marginal likelihood or evidence, i.e.\ the posterior integrated over the full parameter space $\Omega$ allowed by the prior. 

We use an exponential likelihood appropriate for  modeling power spectra in the absence of significant dead time. For a single power spectrum $D_i$ in energy bin $i$ with $M$ powers $\{P_{i,j}\}_{i=1, j=1}^{N,M}$ at each frequency $\nu_j$ we have:

\begin{equation}
\likelihood_i(\pars) = p(D_i \given \pars, \hypers M) = \prod_{i=1}^{M} \frac{1}{\hat{P}_{i,j}(\pars)} \exp{(\frac{P_{i,j}}{\hat{P}_{i,j}(\pars)})} \, , 
\end{equation}

\noindent where $\hat{P}_{i,j}$ denotes the model power at frequency $\nu_j$ in energy bin $i$. We assume  that the observed power spectra in different energy bins are statistically independent; this assumption is valid as long as the photon detection process can be assumed to be largely energy-independent (i.e.\ the probability of detecting a photon of energy $E_1$ does not directly depend on the probability of having previously detected a photon of energy $E_2$) and our energy bins do not overlap. In this case, the joint likelihood for $N$ power spectra becomes:

\begin{equation}
p(\dd \given \pars, \hypers, M) = \prod_{i=1}^{N} \likelihood_i(\pars)  \, .
\end{equation}

Model comparison can now be performed via comparison of the marginal likelihoods. For two models $M_1$ and $M_2$ with priors $p(M_1)$ and $p(M_2)$, and marginal likelihoods $p(\dd \given M_1)$ and $p(\dd \given M_2)$, we can define

\begin{equation}
\label{eqn:bayesfactor}
\frac{p(M_1 \given \dd)}{p(M_2 \given \dd)} = \frac{p(\dd \given M_1)}{p(\dd \given M_2)} \frac{p(M_1)}{p(M_2)} \, .
\end{equation}

\noindent In practice, the posterior  probability distribution is often not analytical, and thus the marginal likelihood difficult or expensive to compute.  However, the expression can be simplified if the models $M_1$ and $M_2$ are nested, that is, if $M_1$ is a special case of $M_2$. Suppose $M_1$ only has parameters $\pars$, and $M_2$ has parameters $\{\pars, \phi\}$, and $M_1$ is a special case of $M_2$ if $\phi = \phi_0$, then the Bayes factor $B_{12} = \frac{p(\dd \given M_1)}{p(\dd \given M_2)}$ reduces to  the Savage-Dickey Density Ratio (SDDR),

\begin{equation}
\label{eqn:sddr}
B_{12} = \frac{p(\phi \given \dd,M _2)}{p(\phi \given \hypers, M_2)}\rvert_{\phi = \phi_0} \, ,
\end{equation}

\noindent where we have integrated $p(\phi \given \dd,M _2) = \int_{\Omega}{p(\pars, \phi \given \dd, \hypers, M_2)d\pars}$ and $\Omega$ denotes the total prior volume spanned by the parameters $\pars$ \citep{Dickey1971}.

In practice, we compute the SDDR by first sampling from $M_2$. We then marginalize over all nuisance parameters (here all parameters in both $M_1$ and $M_2$) and approximate the marginalized posterior distribution $p(\{a_i\}_{i=1}^{N} \given \dd, M_2)$ using a Kernel Density Estimate as implemented in \textit{scikit-learn} \citep{scikit-learn}. This allows us to compute the posterior probability where $a_i = 0 \, \forall i$ as required in Equation \ref{eqn:sddr}.

\subsection{The Model}

We model the power density spectrum in each energy $i$ bin with (1) the Fourier-transformed template  pulsed light curve, scaled appropriately to the photon flux in that energy band from both source and background, (2)  a power law component with parameters $\Gamma_i$ (index) and $A_{\mathrm{PL},i}$ (amplitude) to account for any red noise not accounted for in the pulse profile, (3) a constant $w_i$ to account for the flat white noise component visible in the PSDs at high frequencies. These parameters are shared between $M_1$ and $M_2$. In addition, we consider a scaling factor $A_i$ applied to the pulse profile light curve before Fourier transforming in each energy bin, which accounts for any energy-dependent differences in the pulsed fraction. 

For the simpler model $M_1$, the scaling factor $A_i$ is constant with energy and thus equal for all energy bins, i.e.\ $A_1 = A_2 = \ddots = A_N = C$.
The more flexible model $M_2$ is parametrized as $A_i = C + a_i$, where $C$ is a constant scaling shared by all energy bins, and $a_i$ is an offset in energy bin $i$ from that shared scaling factor. This parametrization is convenient, because the models are nested such that we can recover $M_1$ by setting $a_i = 0 \,\forall i$. Under the assumption that the priors for $a_i$ and $C$ are separable, we can use the SDDR in comparing the two models. Our full set of parameters for $M_1$ is $\pars_1 = \{ \{\Gamma_i\}_{i=1}^{N}, \{A_{\mathrm{PL},i}\}_{i=1}^{N}, \{w_i\}_{i=1}^{N}, C\}$, and for $M_2$ we have $\pars_2 = \{ \{\Gamma_i\}_{i=1}^{N}, \{A_{\mathrm{PL},i}\}_{i=1}^{N}, \{w_i\}_{i=1}^{N}, C, \{a_i\}_{i=1}^{N}\}$.

\subsection{Priors}
\label{sec:priors}

We place largely non-informative priors on our parameters, summarized in Table \ref{tab:priors}. In addition to the individual parameter priors, we impose two additional restrictions. First, for both models we are requiring that all powers in each model power spectrum $\{\hat{P}_{i,j}\}_{i=1, j=1}^{N,M} > 0$, a requirement imposed by the definition of the power spectrum as the square of the real part of the Fourier transform. 

In order to encode the second assumption as a prior, we introduce the model powers at frequency $\nu_j$ in energy bin $i$, $\{\hat{P}_{i,j}\}_{i=1, j=1}^{N,M}$ as latent variables in our model, such that  our posterior for $M_1$ becomes
\begin{align}
&p(\{\hat{P}_{i,j}\}_{i=1, j=1}^{N,M}, \pars_1 \given \dd, \alpha, M_1) =  \\\nonumber 
&\frac{p(\dd \given \{\hat{P}_{i,j}\}_{i=1, j=1}^{N,M}, \pars_1, M_1) p(\pars_1 \given \hypers, M_1) p(\{\hat{P}_{i,j}\}_{i=1, j=1}^{N,M} \given \hypers, M_1)}{p(\dd \given M_1)}  \, .
\end{align}
Because the powers $\{\hat{P}_{i,j}\}_{i=1, j=1}^{N,M}$ are uniquely determined once the parameters $\pars_1$ are known, the prior is a Dirac delta function around the function values, unless any of those function values are negative:
\begin{equation}
  p(\{\hat{P}_{i,j}\}_{i=1, j=1}^{N,M} \given \hypers, M_1) = \begin{cases}
    \delta(P_{i,j} - \hat{P}_{i,j}), & \text{if $\hat{P}_{i,j} > 0 \, \forall i,j$}.\\
    0, & \text{otherwise}.
  \end{cases}
\end{equation}
In practice, this definition of the prior on the latent variables as a delta function allows for straightforward marginalization over these latent variables in order to compute the posterior distribution  over parameters $\pars$ alone. 

\section{Instrument responses}\label{sec:instrument responses}

\begin{figure}
	\centering
		\includegraphics[width=0.45 \textwidth]{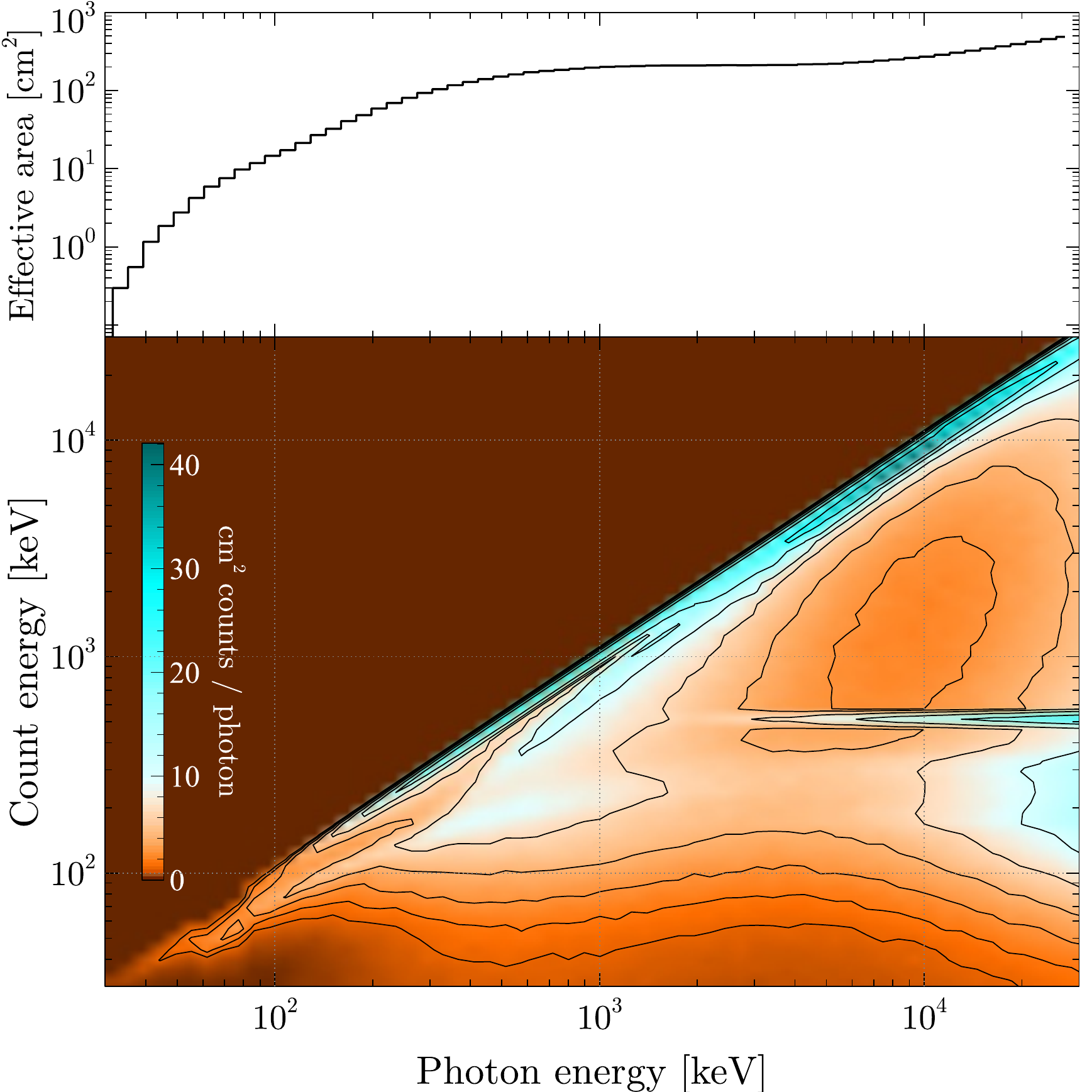}
	\caption{Simulated \emph{RHESSI} response matrix of rear segments at $5^{\circ}$ incidence (bottom panel). The response matrix has been used to generate the fit to the count energy spectrum (Fig.~\ref{fig:spec}) as to obtain an estimate of the incident photon flux. By integrating over the count energy dimension we obtain the effective area of the detector vs incident photon energy (top panel).}\label{fig:RHESSI_rm}
\end{figure}

The \emph{RHESSI} response matrix displayed in Fig.~\ref{fig:RHESSI_rm} represents the response of the detector's rear segments at $5^\circ$ incidence angle and was generated by E. Bellm. We obtain an effective area curve by integrating the response matrix along the axis of the count energies. The effective area indicates the instrument's overall sensitivity to a given photon energy.

\section{Instrumental feature in \emph{RHESSI} data of SGR 1806-20 giant flare}
\label{sec:instrumental feature}

\begin{figure}
	\includegraphics[width= 0.45 \textwidth]{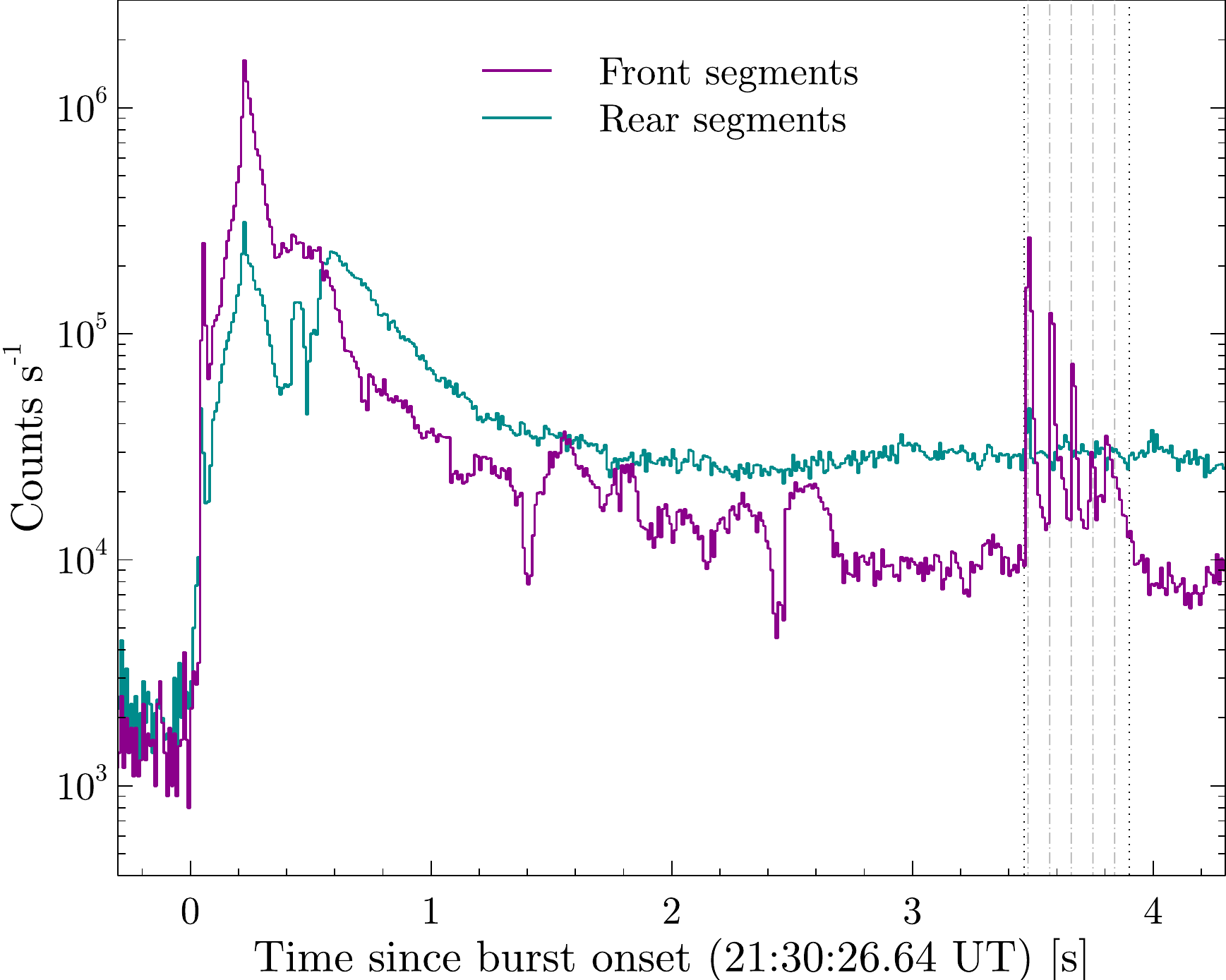}
	\caption{\emph{RHESSI} count rate versus time in the front and rear segments of the detectors. The instrumental feature is apparent at $t-t_0\sim3.465$ s. The dotted lines denote the time-integration interval taken to generate the spectrum (see Fig.~\ref{fig:feature_spec}). The dash-dotted lines roughly indicate the individual peaks and are separated by $\sim0.09$ s. Note that the feature is primarily concentrated to the front segments.} \label{fig:feature_lc}
\end{figure}

\begin{figure}
	\includegraphics[width= 0.45 \textwidth]{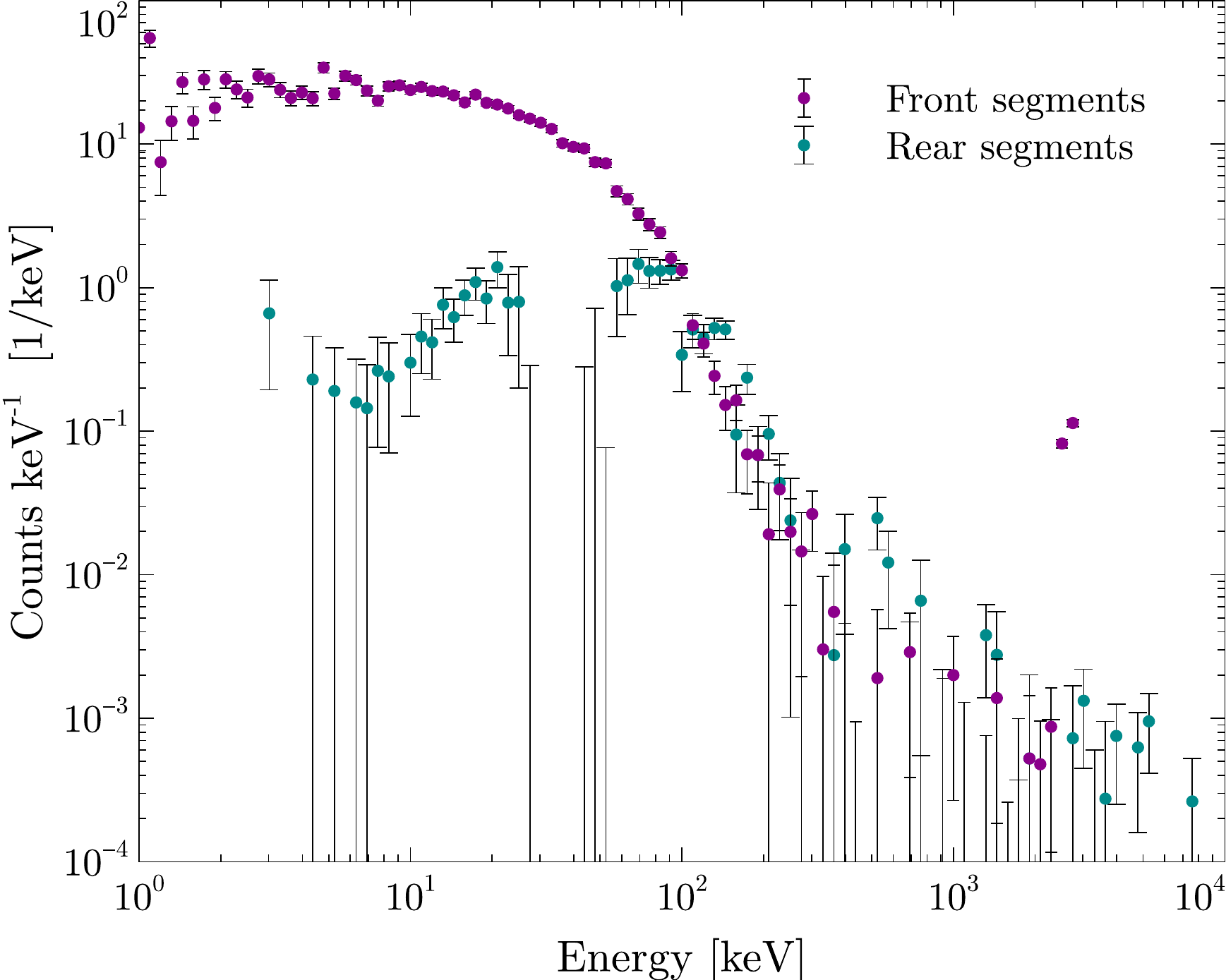}
	\caption{\emph{RHESSI} (background subtracted) spectral energy distribution of the instrumental feature. The magenta (cyan) dots denote the data recorded in the front (rear) segments. Note the presence of a high energy peak at $\sim 2.5-3$ MeV in the front segments.} \label{fig:feature_spec}
\end{figure}

An remarkable feature can be seen in the \emph{RHESSI} light curve at $t_{\rm f}-t_0\sim3.465$ s (see Fig.~\ref{fig:feature_lc}), coinciding with the initial pulse of the pulsating tail of the GF. It is primarily concentrated in the front segments, lasts for $\Delta t_{\rm f}\sim0.435$ s, and comprises a number of spikes, $\sim0.09$ s apart. The background subtracted\footnote{The background spectrum consists of a spectrum integrated over $\Delta t_{\rm f}$ taken just before the appearance of the instrumental feature at $t_{\rm f}-t_0$.} spectrum of the feature is plotted in Fig.~\ref{fig:feature_spec}. The spikes are primarily visible below $\sim 300$ keV and in the range $\sim2.6-3$ MeV in the front segments. 

Following discussion with experts on the \emph{RHESSI} instrumentation (David Smith and Gordon Hurford, private communications), it was determined that these spikes are most likely an instrumental feature caused by the motion of an attenuator that deployed in response to the high count rate. Its motion is induced by putting a current through a wire, which contracts when heated. The heating was initiated at $\sim3.36$ s and the attenuator registered in the `in' position at $\sim3.75$ s. The prompt movement, or rather impact of the attenuator on the stop, can produce microphonic noise in the detectors that may be recorded as an event; generally dominant in the front segments and varying significantly between the individual segments. The separate spikes may be explained by the fact that the attenuator bounced several times after the initial impact before it came to rest -- each strike generating a burst of microphonics.  At the time of writing this effect is not one of the possible sources of lightcurve artifacts described on the \emph{RHESSI} artifact web page\footnote{\emph{RHESSI} artifact web page: \url{http://hesperia.gsfc.nasa.gov/ssw/hessi/doc/guides/lightcurve_artifacts.htm}}. 

This instrumental feature must be taken into account when performing spectral or temporal analysis of the GF in the moments following the initial hard spike in the period $t-t_0 = 3-4$ s. The \emph{RHESSI} high-energy lightcurve (Fig.~\ref{fig:lc_2}) and spectrum (Fig.~\ref{fig:spec}) in this manuscript are composed of data recorded solely in the rear segments, where the presence of the instrumental feature is marginal. Moreover, the behaviour of the pulsating tail as a function of energy, as discussed in Section~\ref{sec:pulsed emission}, is studied in the time range $t - t_0 =5-95$ s, after this instrumental feature has subsided.  


\bsp	
\label{lastpage}
\end{document}